\documentclass[10pt,twocolumn,showpacs,longbibliography,amsmath,amssymb,osajnl,floatfix,superscriptaddress]{revtex4-2}

\usepackage{graphicx}
\usepackage{dcolumn}
\usepackage{color}
\usepackage{txfonts}
\usepackage{microtype}
\usepackage{bm}
\usepackage{ulem}
\usepackage{comment}
\usepackage{amsmath}
\usepackage[export]{adjustbox}

\begin{document}

\title{Laser-driven lepton polarization in the quantum radiation-dominated reflection regime}
\author{Kai-Hong Zhuang}
\author{Yue-Yue Chen}
\email{yue-yue.chen@shnu.edu.cn}
\affiliation{Department of Physics, Shanghai Normal University, Shanghai 200234, China}
\author{Yan-Fei Li}
\email{liyanfei@xjtu.edu.cn}
\affiliation{Department of Nuclear Science and Technology,, Xi’an Jiaotong University, Xi’an 710049, China}
\author{Karen Z. Hatsagortsyan}
\email{k.hatsagortsyan@mpi-hd.mpg.de}
\affiliation{Max-Planck-Institut f\"{u}r Kernphysik, Saupfercheckweg 1, 69117 Heidelberg, Germany}
\author{Christoph H. Keitel}\affiliation{Max-Planck-Institut f\"{u}r Kernphysik, Saupfercheckweg 1, 69117 Heidelberg, Germany}

\date {\today}

\begin{abstract}

Generation of ultrarelativistic polarized leptons during interaction of an ultrarelativistic electron beam with a counterpropagating ultraintense laser pulse is investigated in the quantum radiation-dominated domain. While the symmetry of the laser field tends to average the radiative polarization of leptons to zero, we demonstrate the feasibility of sizable radiative polarization through breaking the symmetry of the process in the reflection regime. After the reflection, the off-axis particles  escape the tightly focused beam with  polarization correlated to the emission angle, while the particles at the beam center are more likely to be captured in the laser field with unmatched polarization and kinetic motion. Meanwhile, polarization along the electric field emerges due to the spin rotation in the transverse plane via precession. In this way, the combined effects of radiative polarization, spin precession and the laser field focusing are shaping the angle-dependent polarization for outgoing leptons. Our spin-resolved Monte Carlo simulations demonstrate an angle-dependent polarization degree up to $\sim$20\% for both  electrons and positrons, with a yield of one pair per seed electron. It provides a new approach for producing polarized high density electron and positron jets at ultraintense laser facilities.


\end{abstract}

\maketitle

\section{introduction}

Polarized electrons and positrons are valuable investigation tools in nuclear and high-energy physics \cite{voutier2014physics,Maas_2004}. They are indispensable for high-energy experiments in the next generation of colliders, which are aiming to search for physics beyond the standard model \cite{moortgat2008polarized}.

There are standard methods for polarization of relativistic electron beams: illuminating a photocathode by circularly polarized light \cite{Pierce_1976}, or by radiative polarization in a storage ring via the Sokolov-Ternov effect \cite{Sokolov_1964,Baier_1967,Derbenev_1973}.
Since antiparticles are not available in the matter, generating intensely polarized positrons is more challenging than polarized electrons. It is well known that $\beta$-decay of specific radioisotopes can produce polarized positrons \cite{zitzewitz1979spin}, however, the quality of the beam is far from practical usage. So far, there are mainly two possible approaches to generate intensely polarized positrons for high-energy physics~\cite{ILC}. One is the so-called, undulator based positron source \cite{alexander2008observation}, producing circularly polarized $\gamma$-photons via electron radiation in a helical undulator, and then converting it to electron-positron pairs in a tungsten target. The proof-of-principle E166 experiment was performed at Stanford Linear Accelerator Center.
The positrons of $2-6\times10^{4}/$pulse are produced with longitudinal polarization above $80\%$ at an energy of about 6 MeV \cite{mikhailichenko2007e166}. Alternatively, $\gamma$-photons can be generated by Compton backscattering of circularly polarized laser  \cite{omori2006efficient}.
In the second method proposed in the Thomas Jefferson National Accelerator Facility, which is termed Polarized Electrons for Polarized Positrons (PEPPo), the polarized positrons are produced using bremsstrahlung radiation of polarized electrons \cite{abbott2016production}.
However, upgrading the intensity of the positron source to meet the requirement of future electron-positron colliders ($2.82\times10^{14} \,\text{s}^{-1}$), is still an extremely challenging task
\cite{scott2011demonstration}.

With the invention of chirped-pulse amplification (CPA) and optical parametric chirped-pulse amplification (OPCPA) laser techniques, the current laser intensity is already able to reach $10^{23}$~W/cm$^2$ \cite{Yoon2021}, and an increase up to $10^{25}$ W/cm$^2$ is expected in next-generation laser facilities \cite{Vulcan,Zou_2015,CORELS,ELI,Exawatt}. Strong lasers have been applied for generating electron-positron jets via laser-solid interaction \cite{Chen_2009,Chen_2010,Chen_2015a,Liang_2015}, as well as via laser-electron beam interaction~\cite{Sarri_2015n}.

Spin effects in nonlinear QED processes in strong laser fields have been investigated theoretically,  in particular, in multiphoton Compton scattering \cite{Kotkin_1998,Panek_2002,Kotkin_2003,ivanov2004complete,Boca_2012,karlovets2011radiative,Krajewska_2013,seipt2018theory}, in Kapitz-Dirac scattering \cite{ahrens2012spin,Dellweg_2016,Dellweg_2017}, in  multiphoton Breit-Wheeler process (pair production) \cite{tsai1993laser+,ivanov2005complete,jansen2016strong}, as well as in multiphoton Bethe-Heilter process in a Coulomb field \cite{di2010polarization}. However, these studies have mostly addressed the processes in a plane-wave field and/or not ultraintense field regimes. In an ultrastrong laser field, the strong-field QED effects are well described within the Local Constant Field Approximation (LCFA) \cite{Baier_1998,Ritus_1985}. An efficient formalism based on LCFA describing polarization effects during a photon emission and pair production in background strong fields has been developed by Baier and Katkov \cite{Baier_1967,Baier_1972, Baier_1998}.

When an unpolarized electron radiates in a strong external field, the electron spin after the emission is preferentially oriented opposite to the magnetic field in the electron rest frame, which is the energetically most favorable state. This effect is termed radiative polarization \cite{Sokolov_1964, Baier_1967}. Similarly,  during pair creation of a linearly polarized/unpolarized photon in a strong external field, the electron (positron) spin favors the opposite (same) direction with respect to the magnetic field (in the electron rest frame), which may lead to the polarization of electron-positron pairs \cite{Baier_1998}. The polarization effect is significantly larger in the case of a pair production process than in that for photon emission. This is because of the larger asymmetry of the  pair production probability between the final spin-up and -down states in a constant field, compared with that of radiation. Available ultrastrong laser fields put forward a desire to exploit intense fields for radiative or pair production polarization of electron (positron) beams. However, its straightforward realization is not possible. In fact, a laser field represents a quite remarkable symmetric field, with the negative and positive half-cycles inducing opposite spin effects. Even in an ultrashort laser field, the field asymmetry is too
weak to allow a significant polarization, e.g., the polarization is less than $8\%$ in a  single cycle plane wave laser pulse discussed in \cite{seipt2018theory}. Therefore, essentially asymmetric laser fields are required to polarize particles either with radiative or pair production polarization. In \cite{del2017spin,del2018electron}  a model asymmetric laser field is devised in the form of a strong rotating electric field, which not surprisingly yields a large electron radiative polarization, reviving the idea to seek radiative (pair production) polarization in more realistic field configurations.

Recently, we have developed a fully polarization resolved Monte Carlo method \cite{li2018ultrarelativistic,li2020polarized,li2020production}  for investigations of spin effects in nonlinear QED processes in ultrastrong focused laser fields, using spin-resolved radiation and pair production probabilities in LCFA, calculated with the Baier-Katkov operator method \cite{Baier_1998,chen2022electron}. Employing this method, we proposed a new concept of producing polarized positrons with strongly focused two-color laser pulses \cite{chen2019polarized}, see also \cite{Seipt_2019}. The asymmetry of the two-color laser field allows for polarization of pairs up to $\sim60\%$  with an efficiency of $10^{-2}$ positrons per electron. Further, with a fine-tuning of the ellipticity of the laser field, the polarized particles arising from different laser cycles can be separated due to spin-dependent radiation reaction, which promotes a further enhancement of polarization \cite{li2018ultrarelativistic,Wan_2019,li_polarimetry_2019}. Positrons density can be further increased to 30 nC by using laser-solid interaction \cite{song2022dense}, or to $10^5–10^6$/bunch by using initially polarized electrons \cite{li2020production}. It also has been proposed to produce polarized leptons via phase-matched radiation reactions between particle momentum and polarization \cite{li2020production,li2022helicity,ma2023production}, {\color{black}followed by postselection technique discussed in detail in Ref. \cite{ma2023production}.}
In another recent development, the electron and photon polarization generated in strong fields in plasma has been employed as a diagnostic tool to monitor transient magnetic fields in plasma \cite{Gong_2021,Gong_2022}, or characterize the ultrarelativistic plasma instabilities \cite{Gong_2023}.


\begin{figure*}[t]
	\includegraphics[width=0.8\textwidth]{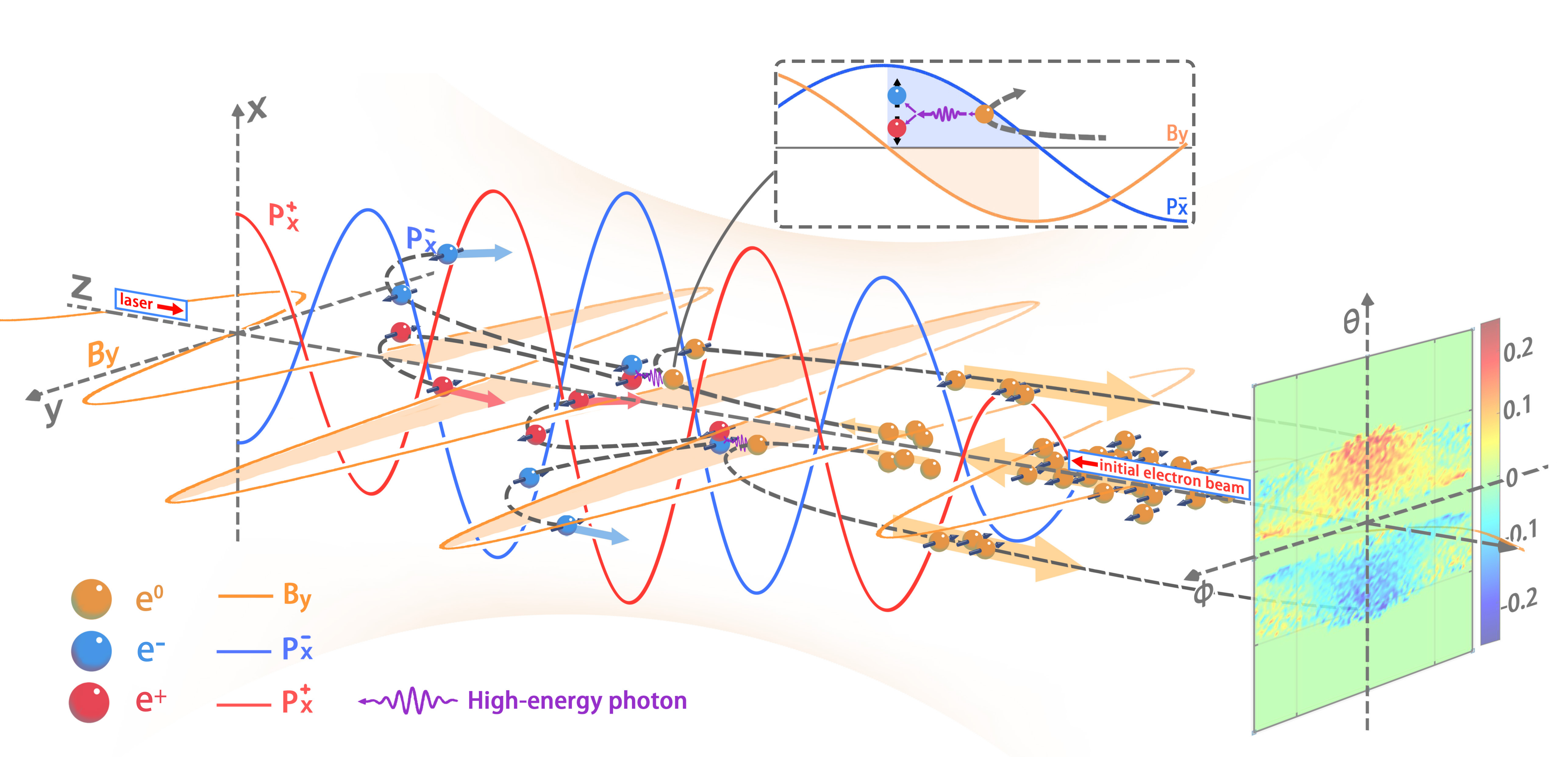}
	\caption{Scheme of laser-based polarized positron beam production. An intense linearly polarized laser pulse head-on collides with an unpolarized relativistic electron beam, resulting in emission of $\gamma$ photons in forward direction, which decay into electron-positron pairs. The produced pairs are reflected in the laser field, and polarized with spin parallel and antiparallel to magnetic field direction for polar angle $\theta>0$ and $\theta<0$, respectively.}
	\label{Fig. scheme}
\end{figure*}

In this paper, we investigate an intense laser beam interaction with a counterpropagating electron beam. Although the applied laser beam is well symmetric, we aim to obtain asymmetric electron-laser field interaction and generate spin polarization of seed electrons and created pairs via invoking the reflection regime of the interaction \cite{di2009strong,li2015attosecond},
see the scheme in Fig.~\ref{Fig. scheme}. Especially advantageous for the creation of asymmetric laser-electron interaction is the applied quantum radiation-dominated regime (QRDR). Our simulation shows that both the reflected seed electrons and pairs are polarized in the intense symmetric field in the reflection regime strengthened with QRDR. The correlation between the cycle-dependent radiative polarization and the reflection angle gives rise to the angle-dependent polarization of particles in a tightly focused symmetric laser field. The physical picture of the interaction and emerging polarization are analyzed in detail.

The QRDR regime is characterized by the following large parameters:  $\chi_e\gtrsim 1$, and  $\alpha a_0\gtrsim 1$   \cite{di2012extremely}. Here, the classical strong field parameter is $a_0=eE_0/m\omega$, and the quantum strong field parameter $\chi_e=|e| \sqrt{|F_{\mu v}p^v|^2}/m^3 $ \cite{Ritus_1985}, with the electron charge $-e$ and mass $m$, the laser field tensor $F_{\mu v}$, the field amplitude $E_0$ and frequency $\omega$, and the electron four-momentum $p^v$ (the relativistic units $\hbar=c=1$ are used).

Our paper is organized as follows. Sec.~\ref{sim} presents the general semiclassical simulation method of spin-resolved laser-electron interaction. The spin-resolved photon emission and pair production probabilities, as well as the definition of the instantaneous quantization axes for simulations, are given in this section.
The results of the numerical simulation are presented in Sec.~\ref{num},
and the polarization for both reflected seed electrons and produced pairs are investigated.  A discussion on how the radiative polarization and the focusing effect of the laser field determine the angle-dependent polarization of seed electrons is presented in Sec.~\ref{num}~A.  The polarization of particles arising from pair creation is analyzed in Sec.~\ref{num}~B. Our conclusion is given in Sec.~\ref{concl}.

\section{Simulation method}\label{sim}

Recently, a semi-classical Monte-Carlo method is developed to describe the electron (positron) spin-resolved  dynamics in nonlinear QED processes in ultrastrong laser fields
\cite{li2018ultrarelativistic,li2020polarized,chen2019polarized,Wan_2019,dai2022photon,li2020production,li2022helicity}. The photon emissions and pair productions are simulated via  Monte Carlo algorithm using spin-resolved quantum probabilities. Between photon emissions and after the pair production, the electrons (positrons) motion in the external field is governed by the Lorentz force, while spin precession is described by Bargmann-Michel-Telegdi (BMT) equation \cite{Bargmann_1959,Walser_2002}.

The spin-resolved photon emission and pair production probabilities are derived with the QED operator method in LCFA \cite{Baier_1998,chen2022electron}.
After summing over the  polarization of emitted photons, we obtain the probability for  emitting a photon with an energy $\omega$ during time step $\Delta t$:
\begin{align}\nonumber
&dW^R(\bm{\zeta},\bm{\zeta}')  =\frac{1}{2} \left( a+\bm{b}\cdot\bm{\zeta}'\right)\\\nonumber
&a =C_0d\omega\left\{ \frac{\varepsilon^{2}+\varepsilon'^{2}}{\varepsilon'\varepsilon}\textrm{K}_{\frac{2}{3}}\left(z_{q}\right)-\intop_{z_q}^{\infty}dx\textrm{K}_{\frac{1}{3}}\left(x\right)-\frac{\omega}{\varepsilon}\bm{\zeta}\mathbf{b}\textrm{K}_{\frac{1}{3}}\left(z_{q}\right)\right\},\\\nonumber
 &\bm{b}=C_0 d\omega\left\{
 \left[2\textrm{K}_{\frac{2}{3}}\left(z_{q}\right)-\intop_{z_{q}}^{\infty}dx\textrm{K}_{\frac{1}{3}}\left(x\right)\right]\bm{\zeta}-\frac{\omega}{\varepsilon'}\textrm{K}_{\frac{1}{3}}\left(z_{q}\right)\mathbf{b}\right.\\\nonumber
 & +\left.\frac{\omega^2}{\varepsilon'\varepsilon}\left[\textrm{K}_{\frac{2}{3}}\left(z_{q}\right)-\intop_{z_{q}}^{\infty}dx\textrm{K}_{\frac{1}{3}}\left(x\right)\right]\left(\bm{\zeta}\mathbf{\hat{v}}\right)\mathbf{\hat{v}}\right\},\\
\end{align}
where $\bm{\zeta}$ and  $\bm{\zeta}'$ are the spin polarization vectors before and after the emission, $\varepsilon$ and $\varepsilon'$ electron energy, $z_q=\frac{2}{3\chi_e}\frac{\omega}{\varepsilon'}$ with $\chi_e$ controlling the magnitude of quantum effects, $C_0=\frac{\alpha}{\sqrt{3}\pi\gamma^2}$, $\mathbf{b}=\mathbf{\hat{v}}\times \bm{s}$ with $\mathbf{\hat{v}}$ and $\bm{s}$ being unit vectors along the direction of electron velocity and acceleration, respectively. The final
polarization vector of the electron resulting from the scattering process itself is
$\bm{\zeta}_f^R=\frac{\bm{b}}{a}$, which is the quantization axis for radiation adopted in our Monte-Carlo simulations.
After each photon emission, the spin of the emitting particle is either parallel or antiparallel to $\bm{n}^R=\bm{\zeta_f^R}$ using the common stochastic algorithm.
If a photon emission event is rejected, one should be aware that the spin of electrons between emissions should be also changed since no-emission probability $W^{NR}$  also has a dependency on initial electron spin \cite{li2020production}:
\begin{align}\nonumber \label{PRB_NR}
&W^{NR}(\bm{\zeta},\bm{\zeta}')=\frac{1}{2}\left(c+\bm{\zeta}'\cdot{\bm d}\right),\\\nonumber
&c=1-\int_0^\varepsilon\widetilde{F}_0d\omega\Delta t,\\
&\bm{d}=\bm{\zeta}\left(1-\int_0^\varepsilon\overline{F}_0d\omega\Delta t\right)+\mathbf{b}C_0\int_0^\varepsilon\frac{\omega}{\varepsilon}\textrm{K}_{\frac{1}{3}}\left(z_{q}\right)d\omega\Delta t,
\end{align}
where
\begin{align*}
\widetilde{F}_0 & =C_0d\omega\left\{ \frac{\varepsilon^{2}+\varepsilon'^{2}}{\varepsilon'\varepsilon}\textrm{K}_{\frac{2}{3}}\left(z_{q}\right)-\intop_{z_q}^{\infty}dx\textrm{K}_{\frac{1}{3}}\left(x\right)-\frac{\omega}{\varepsilon}\bm{\zeta}\mathbf{b}\textrm{K}_{\frac{1}{3}}\left(z_{q}\right)\right\},
\end{align*}
\begin{align*}
\overline{F}_0 & =C_0d\omega\left\{ \frac{\varepsilon^{2}+\varepsilon'^{2}}{\varepsilon'\varepsilon}\textrm{K}_{\frac{2}{3}}\left(z_{q}\right)-\intop_{z_q}^{\infty}dx\textrm{K}_{\frac{1}{3}}\left(x\right)\right\}.
\end{align*}
The no-emission probabilities $W^{NR}_\uparrow$ and $W^{NR}_\downarrow$ are asymmetric with respect to an arbitrary quantization axis $\bm{e}$. Consequently, between photon emissions, the electron spin state along $\bm{e}$ changes to {\color{black} $\zeta^e_f=\frac{W^{NR}(\bm{e})-W^{NR}(\bm{-e})}{W^{NR}(\bm{e})+W^{NR}(\bm{-e})}= \frac{\bm{e}\cdot\bm{d}}{c}$}. Therefore, the final polarization vector without radiation is given by the above expression with $\bm{e}$ taken away, i.e.
$\bm{\zeta_f^{NR}}=\frac{\bm{d}}{c}$.
In Monte Carlo simulations, the spin of an electron between emissions collapses  to one of the two pure states $\pm \bm{n}^{NR}$ with $\bm{n}^{NR}=\bm{\zeta}_f^{NR}/|\bm{\zeta}_f^{NR}|$ using random numbers.
Note that,  the polarization between emissions is physically related to radiative correction \cite{li2022helicity,li2022strong}. Afterwards the spin precession follows BMT equation until the next step.

The emitted photons can further produce electron-positron pairs while propagating in the intense laser field.  After averaging over the polarization of the electrons, the spin-resolved pair production probability of producing an electron with energy $\varepsilon_-$ and positron $\varepsilon_+$ reads \cite{chen2022electron}:

\begin{align}\nonumber\label{Eq.3}
&dW^P(\bm{\xi},\bm{\zeta}^+)=\frac{1}{2}\left( a_++\bm{b}_+\bm{\zeta}^+\right)\\\nonumber
&a_+ =\overline{C}_0d\varepsilon\left\{\int_{z_{p}}^{\infty}dx\textrm{K}_{\frac{1}{3}}\left(x\right)+\frac{\varepsilon^{2}+\varepsilon_{+}^{2}}{\varepsilon\varepsilon_{+}}\textrm{K}_{\frac{2}{3}}\left(z_{p}\right)-\xi_{3}\textrm{K}_{\frac{2}{3}}\left(z_{p}\right)\right\},\\\nonumber
 &\bm{b}_+=\overline{C}_0d\varepsilon\left\{
 \xi_{1}\textrm{K}_{\frac{1}{3}}\left(z_{p}\right)\frac{\omega}{\varepsilon_{+}}\mathbf{s}+\xi_{2}\mathbf{\hat{v}} \left(-\frac{\omega}{\varepsilon}\int_{z_{p}}^{\infty}dx\textrm{K}_{\frac{1}{3}}\left(x\right)\right.\right.\\\nonumber
 & \left.\left.+\frac{\varepsilon_{+}^{2}-\varepsilon^{2}}{\varepsilon\varepsilon_{+}}\textrm{K}_{\frac{2}{3}}\left(z_{p}\right)\right)+\left(\frac{\omega}{\varepsilon}-\xi_{3}\frac{\omega}{\varepsilon_{+}}\right)\mathbf{b}\textrm{K}_{\frac{1}{3}}\left(z_{p}\right)\right\}.\\
\end{align}
Here $\overline{C}_0=\frac{\alpha m^{2}}{\sqrt{3}\pi\omega^{2}}$, $z_p=\frac{2}{3\chi_\gamma}\frac{\omega^2}{\varepsilon_+\varepsilon_-}$,
$\bm{\zeta}^+$ and $\bm{\xi}=\left(\xi_1,\xi_2,\xi_3\right)$ are the spin polarization vectors of a produced positron and stokes parameters of a parent photon, respectively. The quantum parameter for pair production is $\chi_\gamma=|e| \sqrt{(F_{\mu\nu}k^\nu)^2}/m^3$, with four-momentum $k=(\omega,\bm{k})$ of the incoming $\gamma$-photon. The instantaneous spin quantization axis for pair production is along $\bm{\zeta}_f^+=\frac{\bm{b}_+}{a_+}$. If the photon is linearly polarized with $\bm{\xi}=\left(0,0,\xi_3\right)$ or unpolarized,  $\bm{\zeta}_f^+$ becomes the magnetic field direction in the frame of the center-of-mass of the produced pairs. The polarization of the produced pairs is decided by probability  of Eq.~(\ref{Eq.3}) within the Monte-Carlo algorithm. After the pair  is produced, the parent $\gamma$-photon is removed from the simulation, and the created particles can further emit photons and again produce pairs, which could lead to a cascade in the case of  large $\chi_e$. In a  similar manner, photon polarization has also been included in the simulation (see more details in \cite{li2020polarized,dai2022photon,li2020production}).

\section{Results and analysis}\label{num}

\begin{figure}[b]
	\includegraphics[width=0.5\textwidth]{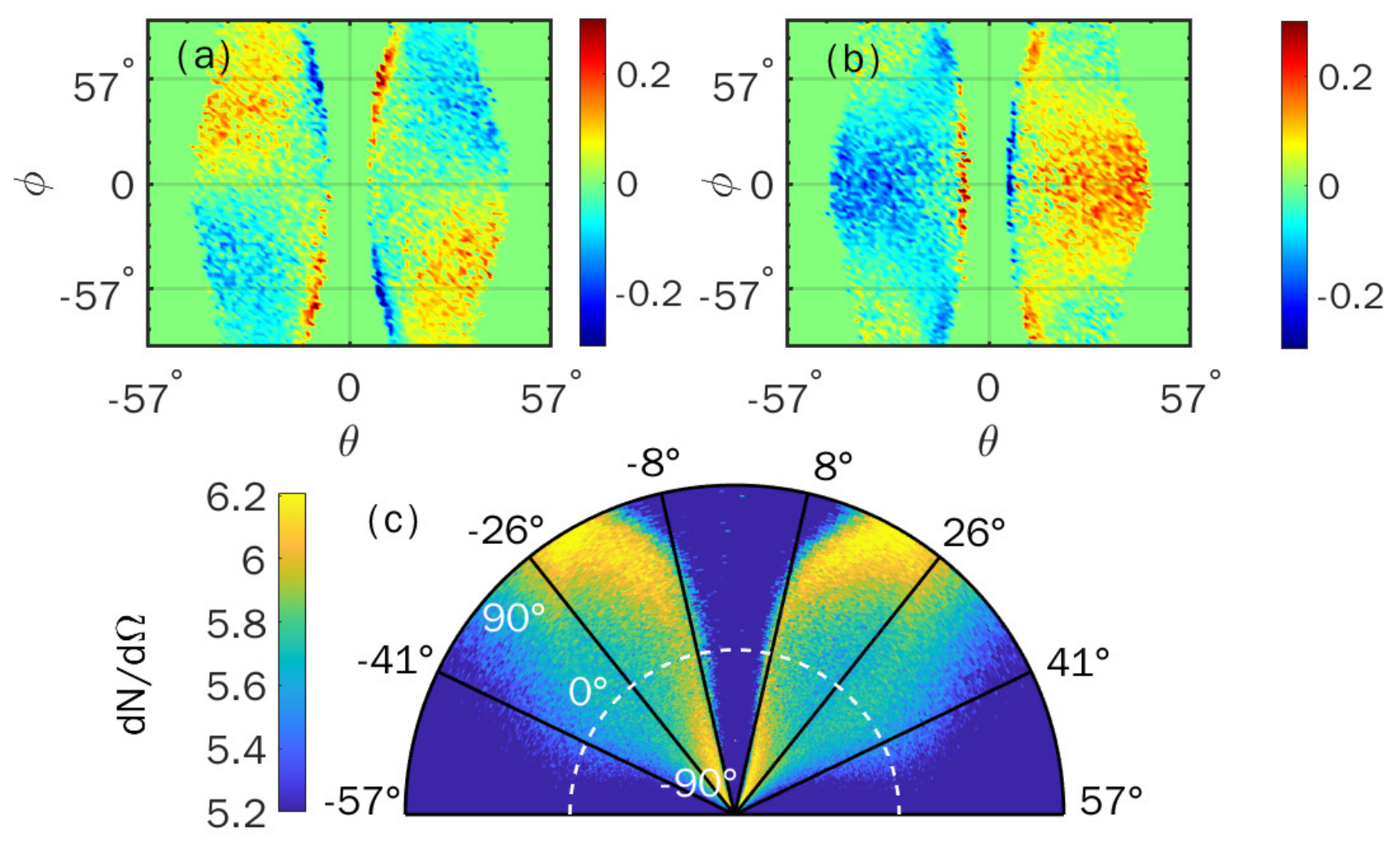}
	\caption{The averaged polarization distribution  (a) $\zeta_x$, (b) $\zeta_y$, and (c) angular distribution d$^2N$/d$\Omega$ vs the polar angle $\theta$ (degree, black scale) and azimuthal angle $\phi$ (degree, white scale) for seed electrons after the interaction at $t=20T$.}
	\label{Fig. agl}
\end{figure}

{\color{black} In this section, we present the results of our Monte Carlo simulations.} We consider an ultra-relativistic electron beam with energy $\varepsilon_0=1.3$ GeV colliding with a tightly-focused linearly polarized strong laser field with intensity $a_0=760$ ($I\sim10^{24}$W/cm$^2$), see the scheme in Fig. \ref{Fig. scheme}. {\color{black}In the simulation the electrons move in a given laser field.}
The laser pulse duration is $\tau_p=3T$ with T being the laser period, beam waist $w_0=2\lambda_0$ {\color{black}and wavelength $\lambda_0=1\mu$m}. The electron beam in the simulation consists of $\sim 10^6$ electrons that have a uniform distribution in the longitudinal direction and a Gaussian distribution in the transverse direction. The length of the electron beam is $L_e=1.5\lambda_0$, the radius $r_e=\lambda_0$. The angular spreading of the electron beam is $\Delta \theta=1$ mrad, and the energy spreading $\Delta\varepsilon/\varepsilon=0.02$, which are typical for laser wakefield acceleration of electrons \cite{esarey2009physics}. The parameters are chosen such that the interaction is in the QRDR regime,  when $\chi_e\approx 2 {\color{black}a_0}(\omega/m)\gamma>1$ and $\alpha {\color{black}a_0}\chi_e>1$, with the electron effective energy in the laser field $\varepsilon =m\gamma$. In addition, the electrons experience a large energy loss, positrons are produced with high density, and reflection occurs as ${\color{black}a_0}/2>\gamma$ which breaks the symmetry of the interaction. Note that due to large radiation losses, the electron energy in the laser field $\varepsilon$ is significantly smaller than $\varepsilon_0$. The specific choice of the laser and electron parameters is to allow the electrons to be reflected near the laser pulse peak.

\begin{figure}[b]
	\includegraphics[width=0.5\textwidth]{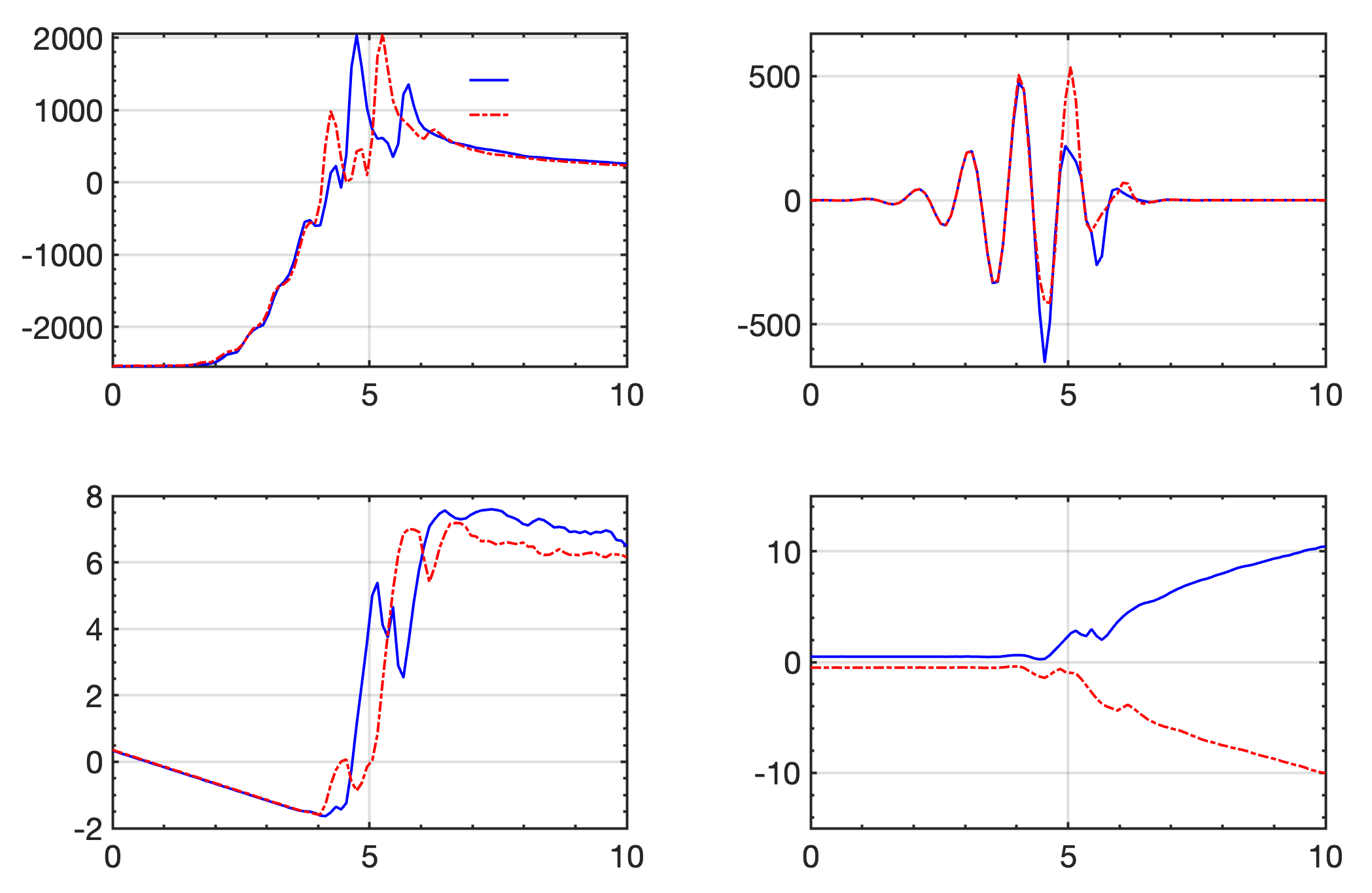}
	\caption{ The electron dynamics vs the laser phase $\varphi$: (a) $\overline{p}_z$ normalized to $m$,  (b) magnetic field $\overline{B}_y$  normalized to $m\omega/|e|$, (c) trajectory $\overline{z}$ and (d) $\overline{x},$ normalized to $\lambda_0$. The overline indicates averaging over electrons within  $(\theta,\phi)\in[-20^\circ,20^\circ]$ for the final momentum $p_x^f>0$ (blue solid line) and $p_x^f<0$ (red dashed line). }
\begin{picture}(300,-30)
  \put(27,225){\small (a)}
  \put(153,225){\small(b)}
  \put(27,135){\small (c)}
  \put(153,135){\small(d)}
  \put(98,228){\tiny $p_x^f>0$}
  \put(98,218){\tiny $p_x^f<0$}
  \put(-5,203){\rotatebox{90}{\small $\overline{p}_z$}}
  \put(0,116){\rotatebox{90}{\small $\overline{z}$}}
  \put(133,116){\rotatebox{90}{\small $\overline{x}$}}
  \put(130,203){\rotatebox{90}{\small $\overline{B}_y$}}
   \put(61,72){{\small $\varphi/(2\pi)$}}
  \put(188,72){{\small $\varphi/(2\pi)$}}
  \end{picture}
	\label{Fig. evl}
\end{figure}
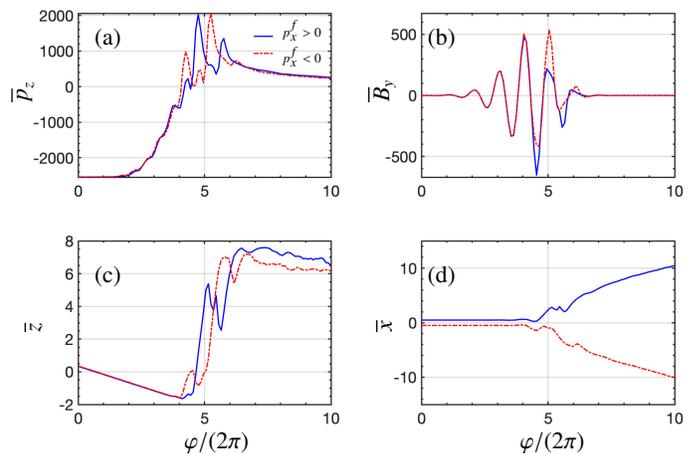

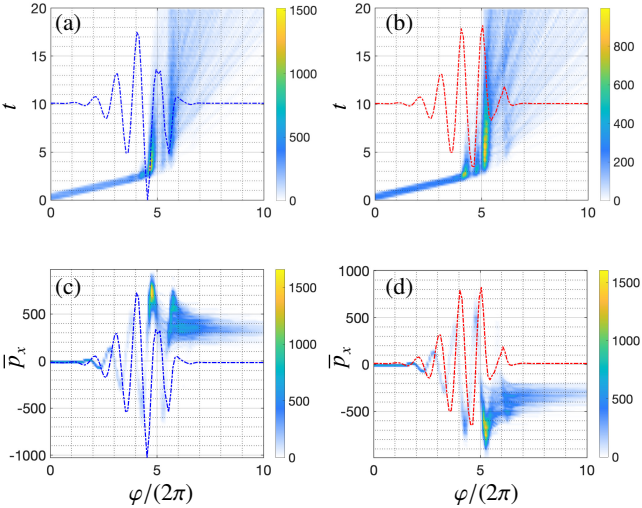
\begin{figure}[b]
	\includegraphics[width=0.5\textwidth,right]{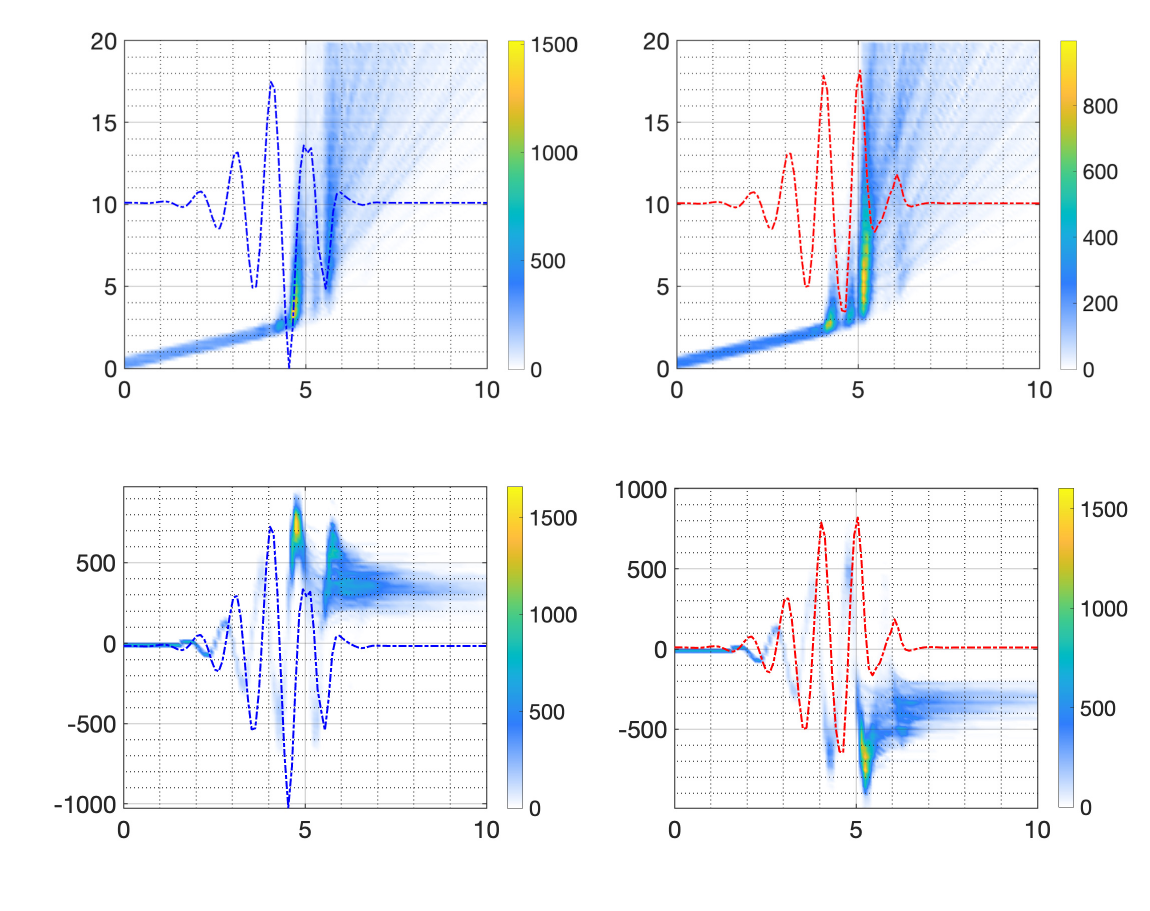}
	\caption{ (Upper row) The number density $\text{d}^2N/(\text{d}t\text{d}\varphi)$ versus evolution time $t/T$ and phase $\varphi/2\pi$, and (bottom row) the number density $\text{d}^2N/(\text{d}p_x\text{d}\varphi)$ versus $p_x$ and $\varphi/2\pi$: for electrons with final momentum $p_x^f>0$ (left column) and $p_x^f<0$ (right column). The electrons are within $(\theta,\phi)\in[-20^\circ,20^\circ]$. The superimposed dot-dashed lines are the corresponding magnetic fields.}
\begin{picture}(300,-30)
  \put(20,270){\small (a)}
  \put(145,270){\small(b)}
  \put(20,170){\small (c)}
  \put(145,170){\small(d)}
  \put(0,240){\rotatebox{90}{\small $t$}}
  \put(0,142){\rotatebox{90}{\small $\overline{p}_x$}}
  \put(125,240){\rotatebox{90}{\small $t$}}
  \put(123,142){\rotatebox{90}{\small $\overline{p}_x$}}
   \put(48,92){{\small $\varphi/(2\pi)$}}
  \put(171,92){{\small $\varphi/(2\pi)$}}
  \end{picture}
	\label{Fig. tpxphi}
\end{figure}

\subsection{Radiative polarization of seed electrons}

The angular distribution and polarization of seed electrons after the interaction are presented in Fig.~\ref{Fig. agl}. The strong interaction induces  a large number of photon emissions, more than one photon per laser cycle $N_\gamma\sim\alpha {\color{black}a_0}\gtrsim 1$ \cite{di2009strong}. The electrons with considerable energy loss could be reflected when the external field exerts a Lorentz force along the laser propagation direction. In the considered reflection regime, after the interaction, the seed electrons move in the reflection direction $|\theta|=|\cos^{-1}(p_z/\gamma)| < 90^0$ ($\theta$ is the polar angle between the particle momentum and the laser propagation direction), as shown in Fig.~\ref{Fig. agl} (c).
Interestingly, the seed electrons are transversely polarized with the highest polarization degree around $20\%$. Along the laser polarization direction (x-axis), the electrons are polarized with central symmetry, see Fig. \ref{Fig. agl} (a). For $\theta>20^\circ$ ($\theta<-20^\circ$),  $\zeta_x$ is positive at $\phi<0$ ($\phi>0$) and negative at $\phi>0$ ($\phi<0$).  Meanwhile, polarization along the magnetic field direction  (y-axis) is opposite with regard to $\theta=0$, see Fig. \ref{Fig. agl} (b).
The electrons with $\theta<0$ has negative $\zeta_y$ while that with $\theta>0$ has positive $\zeta_y$.
It is worth noting that the polarization direction in the small angle region ($|\theta|\lesssim20^\circ$) is opposite to that in the large angle region ($|\theta|\gtrsim20^\circ$).

\subsubsection{Large angle electrons: Polarization along the magnetic field}

The investigation of the angle-dependent polarization, we start with the analysis of the electron dynamics. We calculate the average  momentum evolution and the trajectories for electrons with $\theta>20^\circ$ and $\theta<-20^\circ$, respectively [see Fig. \ref{Fig. evl}]. On average, the electrons with final momentum $p_x^f>0$ and $p_x^f<0$ are initially distributed with $\overline{x}(0)=\pm\lambda_0/2$, respectively. Generally, the electrons are reflected due to the damping in $z$-motion induced by radiation reaction and further acceleration in the laser propagation direction induced by the Lorentz force. Defining the reflection point in terms of the laser phase by $\overline{p}_z=0$,
the   electrons initially distributed at $\bar{x}(0)=\pm\lambda_0/2$ are reflected at the laser phase $\varphi/2\pi=4.2$ and $\varphi/2\pi=4.1$, respectively [see Fig. \ref{Fig. evl} (a)].

In the case of $a_0\gg1$, the reflected electrons become ultra-relativistic along the laser propagation direction within
the laser quarter-cycle due to the Lorentz force effect. Because of the ultrarelativistic drift in the laser propagation direction, the electron dephasing time with respect to the laser field is much longer than the laser period.
At the same time, the electrons are transversely pushed away from the beam axis by the ponderomotive potential of the tightly focused laser field. As illustrated in Figs.~\ref{Fig. evl} and \ref{Fig. tpxphi}, the electrons initially distributed at $\overline{x}(0)=\lambda_0/2$ are accelerated to $p_z\sim 2\times 10^3$ in the acceleration cycle $4.55<\varphi/2\pi<4.8$ [Fig. \ref{Fig. evl} (a)] and stay in this cycle for a long time until slowly drift to the deceleration cycle $4.8<\varphi/2\pi<5.05$ [Fig.\ref{Fig. tpxphi} (a)].  During the phase-matched motion at $4.55<\varphi/2\pi<5.05$, the transverse coordinate  $\overline{x}$ increases up to $2.6\lambda_0$ [Fig. \ref{Fig. evl} (d)] because $\overline{p}_x$ remains positive [Fig. \ref{Fig. tpxphi} (c)]. As the employed laser field is short, during the phase-matched motion, the electrons travel the half of the Rayleigh length of the laser beam [$\overline{z}$ reaches to $5\lambda_0$, see Fig. \ref{Fig. evl} (c)], where the field intensity is significantly reduced.  After the further drift to the next half-cycle $5.05<\varphi/2\pi<5.55$, the laser field attempts to pull the electrons back towards the beam center. However, the rather weak field  [Fig. \ref{Fig. evl} (b)] could not flip the sign of $\overline{p}_x$. Eventually, the electrons are scattered out of the beam in the transverse direction [see Figs.~\ref{Fig. evl} (c),(d) where $\overline{x}$ is increasing, while $\overline{z}$ decreasing at $\varphi/2\pi >6$]. In our stochastic simulation, some electrons directly fly out of the laser beam at the half-cycle $4.55<\varphi/2\pi<5.05$. Those are the electrons with linear time dependence on the phase until  $20T$ [Fig.~\ref{Fig. tpxphi}~(a)],  and having a constant $p_x$ after $\varphi/2\pi=5.05$ [Fig. \ref{Fig. tpxphi} (c)]. However, the most of electrons are scattered out of the beam at $5.55<\varphi/2\pi<6.05$ [Fig. \ref{Fig. tpxphi} (a) and (c)].


Thus, the electron dynamics is mostly determined by the half-cycles near the pulse peak, which could exert a strong transverse acceleration to push the electrons out of the laser beam.
For electrons distributed with $x(0)>0$, $p_x^f$ is mainly affected by the negative half-cycle peaked at $\varphi/2\pi=5.55$, which catches the electrons in the acceleration phase of $p_z$ and provides strong transverse acceleration along $x>0$, pushing the electrons further away from the beam center. Due to the features of the focused laser beam, the effects of the next positive half-cycle on the electron's motion cannot cancel out that from the previous negative half-cycle. Similarly, the dynamics of electrons distributed with $x(0)<0$ is mainly determined by the positive half-cycle peaked at $\varphi/2\pi=5.05$, which catches the electrons in the acceleration phase of $p_z$ [Fig. \ref{Fig. tpxphi} (b)] and exerts a strong negative transverse acceleration [Fig. \ref{Fig. tpxphi} (d)].

\begin{figure}[t]
	\includegraphics[width=0.5\textwidth]{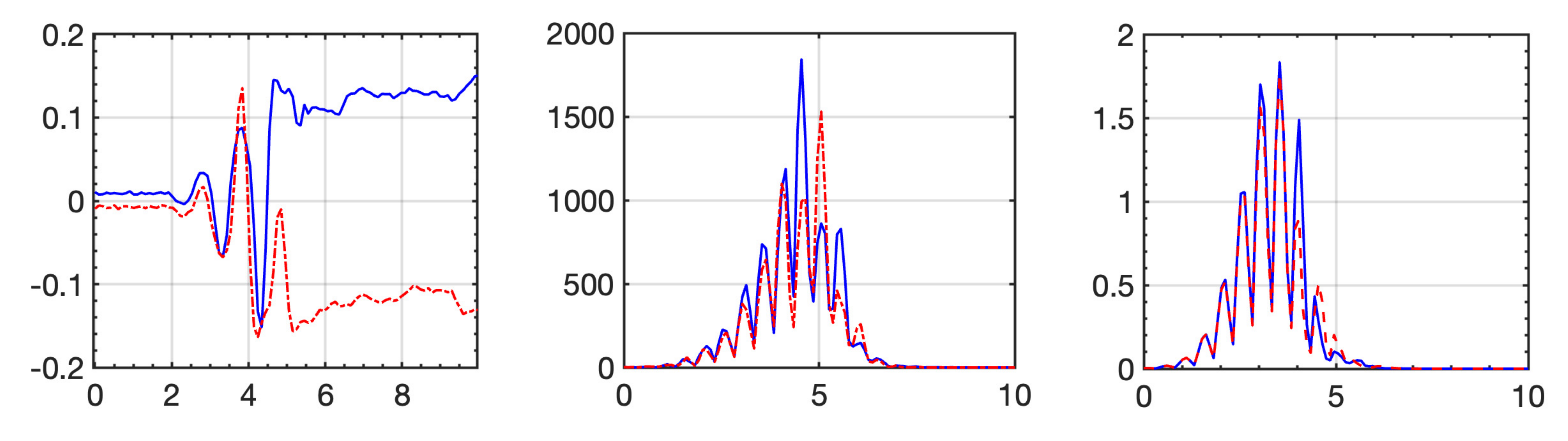}
	\caption{The evolution of $y-$component polarization $\overline{\zeta}_y$ (a), emitted photon number $\overline{n}_\gamma$ (b) and  $\overline{\chi}_e$ (c) averaged over electrons with $(\theta,\phi)\in[-20^\circ,20^\circ]$: for final momentum $p_x^f>0$ (blue solid line) and $p_x^f<0$ (red dashed line).}
	\label{Fig. pol}
		\begin{picture}(300,-30)
  \put(20,125){\small (a)}
  \put(103,125){\small(b)}
  \put(190,125){\small (c)}
  \put(-4,102){\rotatebox{90}{\small $\overline{\zeta}_y$}}
  \put(80,102){\rotatebox{90}{\small $\overline{n}_\gamma$}}
  \put(171,102){\rotatebox{90}{\small $\overline{\chi}_e$}}
  \put(36,62){{\small $\varphi/(2\pi)$}}
  \put(123,62){{\small $\varphi/(2\pi)$}}
  \put(208,62){{\small $\varphi/(2\pi)$}}
  \end{picture}
\end{figure}
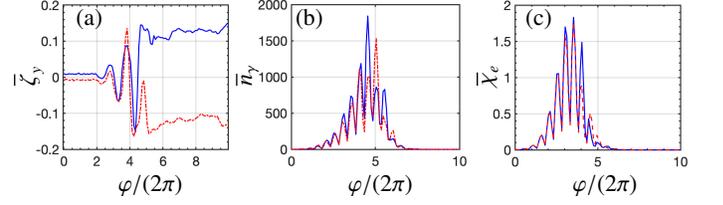

With the information on the electron dynamics at hand, we proceed to interpret the correlation of momentum and polarization [see Fig. \ref{Fig. pol}]. After the reflection at $\varphi/2\pi=4.2$,  the electrons with $\overline{x}(0)>0$ are accelerated and are phase-matched by the main negative half-cycle peaked at $\varphi/2\pi=4.55$, where the electron's average quantum strong-field parameter $\overline{\chi}_e$ is large [Fig.~\ref{Fig. pol}(c)], and the electrons emit a considerable amount of photons [Fig. \ref{Fig. pol} (b)] with a significant recoil due to the large $\overline{\chi}_e$. This leads to polarization of electrons antiparallel with the magnetic field direction, i.e. $B_y<0$, $\zeta_y>0$. After the main negative peak, $\chi_e$ (spin effects) and emission times are both suppressed due to copropagation geometry and weakened fields [Fig. \ref{Fig. pol} (c)], consequently $\overline{\zeta}_y$ is relatively constant after $\varphi/2\pi=4.55$ [Fig. \ref{Fig. pol} (a)]. Similarly, the electrons with $\overline{x}(0)<0$ are accelerated by the main positive half-cycle peaked at $\varphi/2\pi=5.05$, where the electrons are radiatively polarized with $\zeta_y<0$ [see Fig. \ref{Fig. pol} (b)]. Therefore, we conclude that the polarization, similar to the electron motion, is also mostly determined by the main acceleration half-cycle, leading to the correlation of $\zeta_y$ and $p_x$.

\subsubsection{Large angle electrons: Polarization along the electric field}\label{sx}

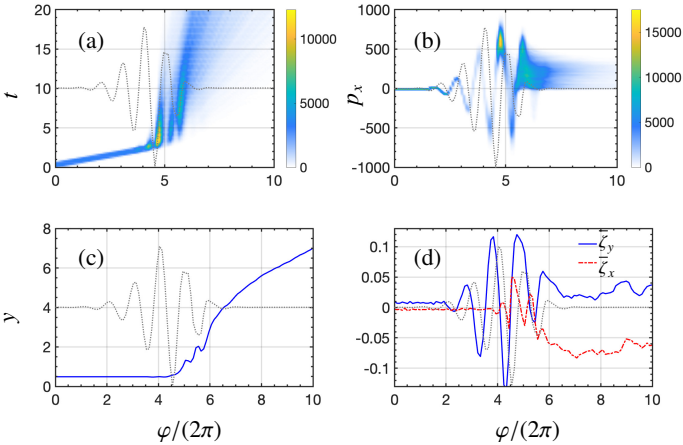
\begin{figure}[b]
	\includegraphics[width=0.5\textwidth]{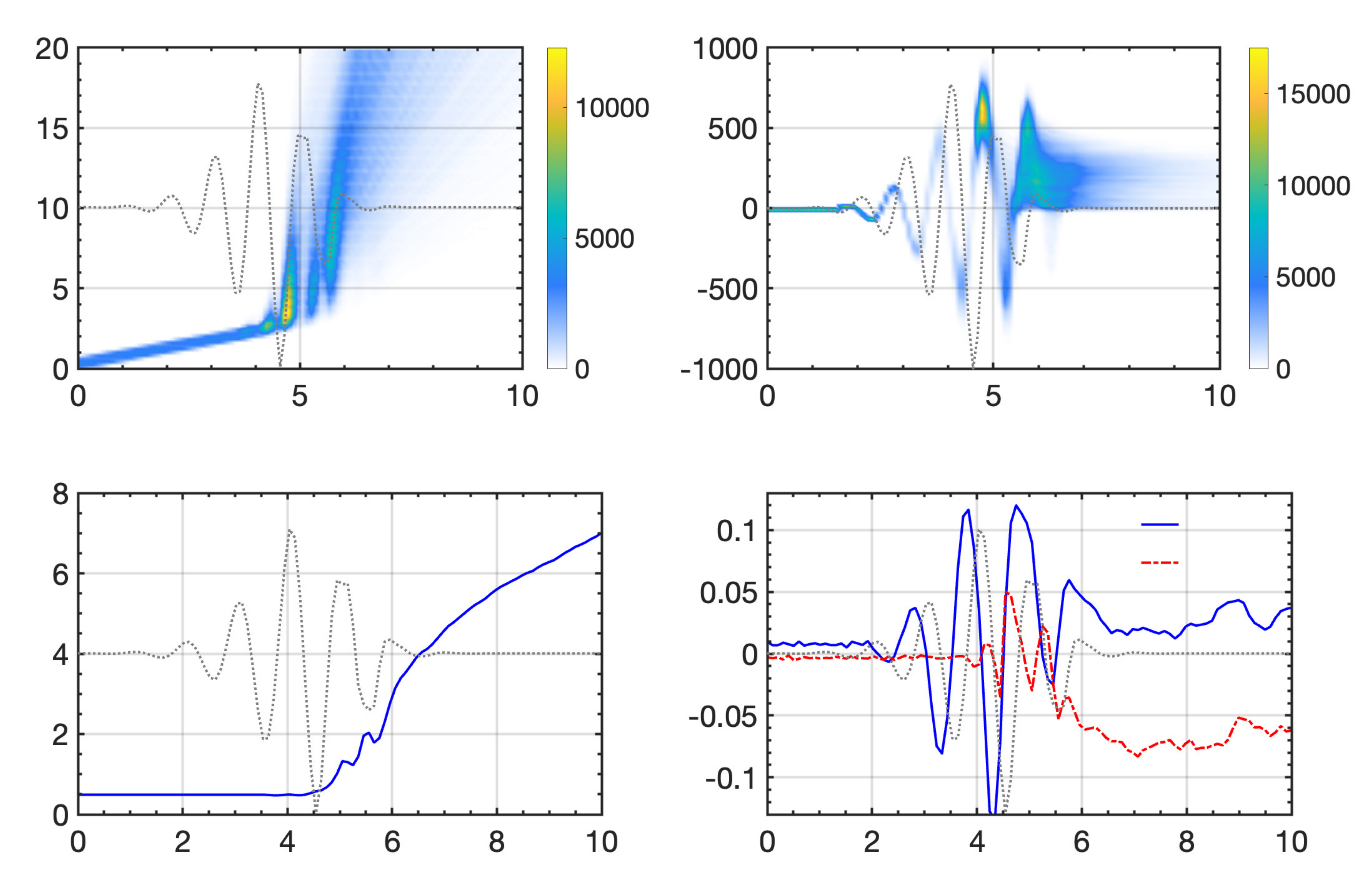}
	  \begin{picture}(300,-30)
  \put(23,152){\small (a)}
  \put(150,152){\small(b)}
  \put(23,70){\small (c)}
  \put(150,70){\small(d)}
  \put(-5,133){\rotatebox{90}{\small $t$}}
  \put(125,130){\rotatebox{90}{\small $p_x$}}
  \put(-5,48){\rotatebox{90}{\small $y$}}
  \put(220,77){{\scriptsize $\overline{\zeta}_y$}}
  \put(220,67){{\scriptsize $\overline{\zeta}_x$}}
   \put(53,5){{\small $\varphi/(2\pi)$}}
  \put(181,5){{\small $\varphi/(2\pi)$}}
 \end{picture}
	\caption{(a) The number density $\text{d}^2N/(\text{d}t\text{d}\varphi)$ vs evolution time $t/T$ and phase $\varphi/2\pi$. (b) The number density $\text{d}^2N/(\text{d}p_x\text{d}\varphi)$ vs $p_x$ and $\varphi/2\pi$. (c) The evolution of $y/\lambda_0$  vs laser phase $\varphi/2\pi$. (d) Average polarization $\overline{\zeta}_x$ (blue solid line) and $\overline{\zeta}_y$ (red dashed line) vs laser phase $\varphi/2\pi$. The electrons are within $\theta>20^\circ$ and $\phi>20^\circ$. The superimposed dot-dashed lines in (a) and (b) are the corresponding magnetic fields. }
	\label{Fig. sx}
\end{figure}

The spin precession also plays a role in  electron spin dynamics, leading to rotation of the polarization vector along velocity direction with a frequency
\begin{eqnarray}\label{precession}
\frac{d\zeta_x(\varphi)}{d\varphi} &=&-\omega_s(\varphi)\zeta_y(\varphi)\\
  \omega_s(\varphi)&\approx&\frac{|e|\gamma}{m k\cdot p}\left[\left(\frac{g}{2}-1\right)\left(1-\frac{\gamma}{\gamma+1}v_{z}\right)+\frac{1}{\gamma+1}\right]v_{y}E_{x}(\varphi).\nonumber
\end{eqnarray}
Before the arrival of the laser pulse peak, the spin precession is insignificant due to the smallness of $v_y$, while becoming notable near the pulse peak, especially for electrons moving at large $(\theta, \phi)$, due to the enhancement of ponderomotive force associated with finite beam size. Take the electrons within $\theta>0$ and $\varphi>20^\circ$ for an example, see Fig. \ref{Fig. sx}.  These electrons are initially distributed with $\overline{y}(0)>0$ and move in the $x-z$ plane before they meet the laser peak at $\varphi/2\pi=4.55$ [Fig. \ref{Fig. sx}(c)]. When the laser peak arrives, the electrons are accelerated to $p_y>0$ because of the ponderomotive force $F_y\propto y>0$. Consequently, the spin precession starts to take effect inducing a transverse polarization along the electric field $\zeta_x$ via Eq.~\ref{precession}
[Fig. \ref{Fig. sx} (d)]. While the spin dynamics of $\zeta_y$ is still determined by radiative polarization until the electrons run away from the beam center after $\varphi/2\pi\approx5.55$ [Fig. \ref{Fig. sx} (a) and (b)].
Assuming the electrons move in a monochromatic plane wave $E(\varphi)=E_0\cos\varphi$,
the radiative polarization gives rise to a transverse polarization $\zeta_y\propto -\sin\varphi$, while the spin precession induces a rotation of the spin to $\zeta_x\propto -\cos(2\varphi)/4$, which qualitatively is in accordance with the spin dynamics shown in Fig. \ref{Fig. sx} (d). As electrons are escaping the laser beam at the negative half-cycle peaked at $\varphi/2\pi=5.55$, they gain positive $\zeta_y$ and negative $\zeta_x$, as shown in Fig. \ref{Fig. agl} (a) and (b).
In contrast, the electrons with small $\phi$ are free from spin precession even after the main peak arrives, since the ponderomotive force along $y$ ($F_y\propto y$) is negligible along $x$-axis ($\varphi=0^\circ$). Therefore, the  electrons distributed at $\varphi\sim0$ have negligible $\zeta_x$ and a maximum for $\zeta_y$,  see Figs.~\ref{Fig. agl}~(a) and (b).


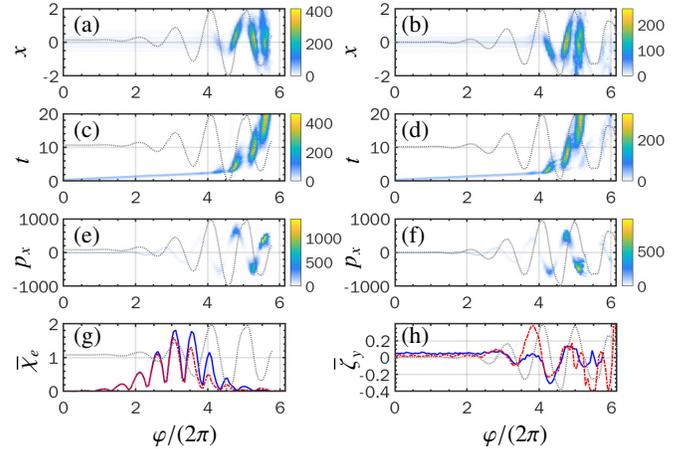
\begin{figure}[b]
	\includegraphics[width=0.5\textwidth]{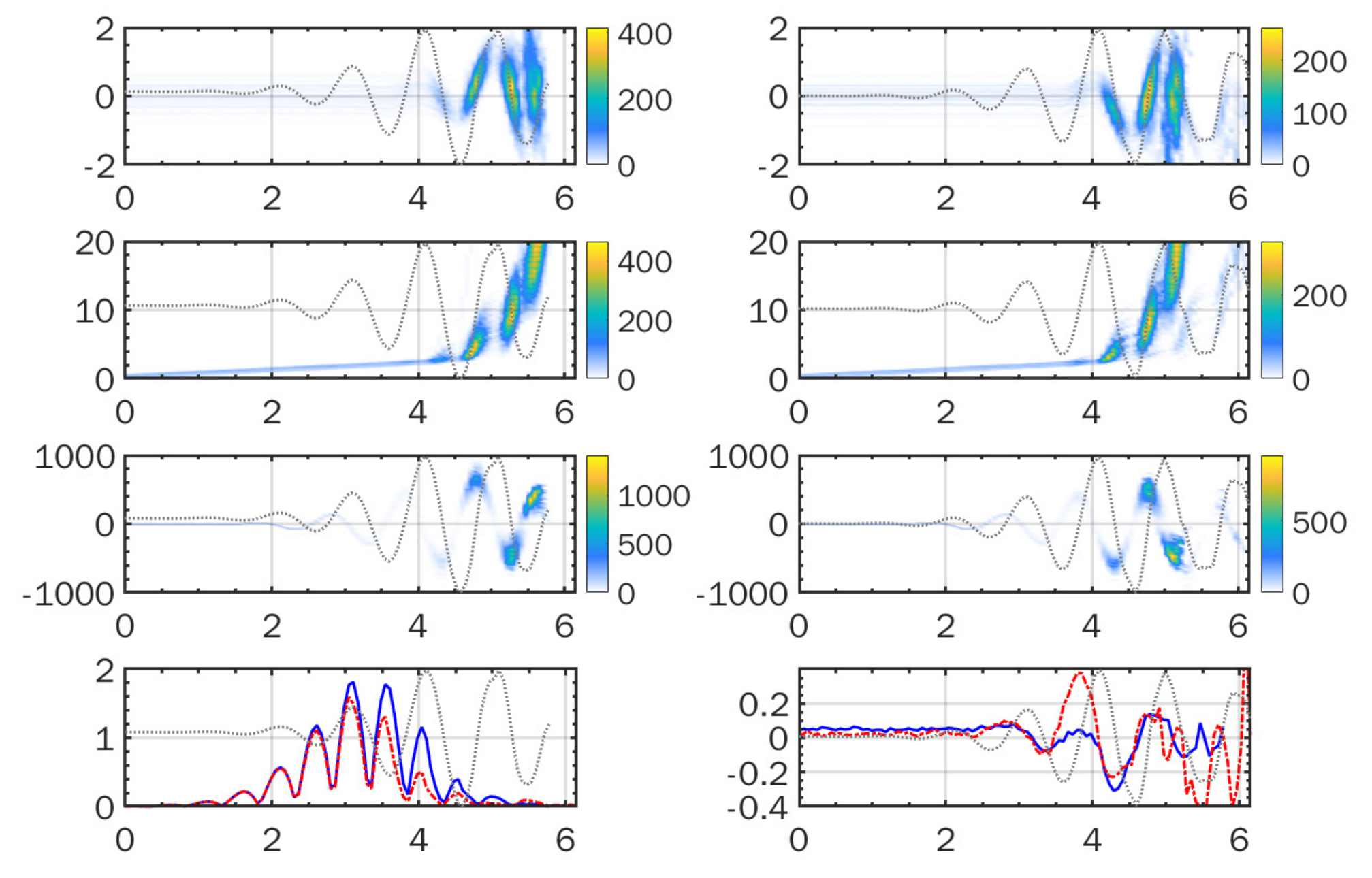}
	  \begin{picture}(300,-30)
  \put(27,160){\small (a)}
  \put(151,160){\small(b)}
    \put(27,120){\small (c)}
  \put(151,120){\small(d)}
  \put(27,80){\small (e)}
  \put(151,80){\small(f)}
  \put(27,42){\small (g)}
  \put(151,42){\small(h)}
  \put(5,150){\rotatebox{90}{\small $x$}}
  \put(130,150){\rotatebox{90}{\small $x$}}
  \put(5,110){\rotatebox{90}{\small $t$}}
  \put(130,110){\rotatebox{90}{\small $t$}}
  \put(5,70){\rotatebox{90}{\small $p_x$}}
  \put(130,70){\rotatebox{90}{\small $p_x$}}
  \put(5,32){\rotatebox{90}{\small $\overline{\chi}_e$}}
  \put(125,32){\rotatebox{90}{\small $\overline{\zeta}_y$}}
   \put(57,5){{\small $\varphi/(2\pi)$}}
  \put(182,5){{\small $\varphi/(2\pi)$}}
 \end{picture}
	\caption{ (Top row) The trajectory $x$, (second row) evolution time $t$, and (third row) momentum evolution $p_x$ vs the laser phase $\varphi$, for electrons within $\theta\in[-10^\circ, 10^\circ]$ and $\phi\in[-20^\circ, 20^\circ]$; for electrons with final momentum $p_x^f>0$  (left column), and $p_x^f<0$ (right column); (g) $\overline{\chi}_e$  and  (h) $\overline{\zeta}_y$  vs the laser phase $\varphi$ for electrons with final momentum $p_x^f>0$  (blue solid line) and $p_x^f<0$ (red dashed line).}
	\label{Fig. small}
\end{figure}

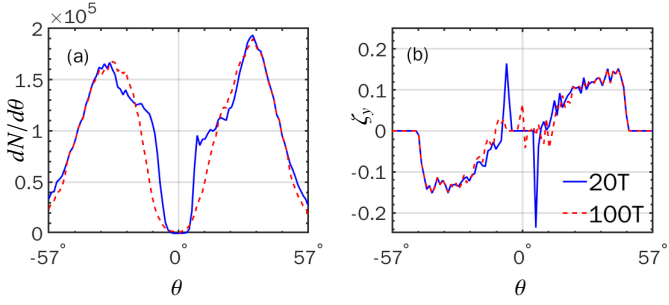
\begin{figure}[]
	\includegraphics[width=0.5\textwidth]{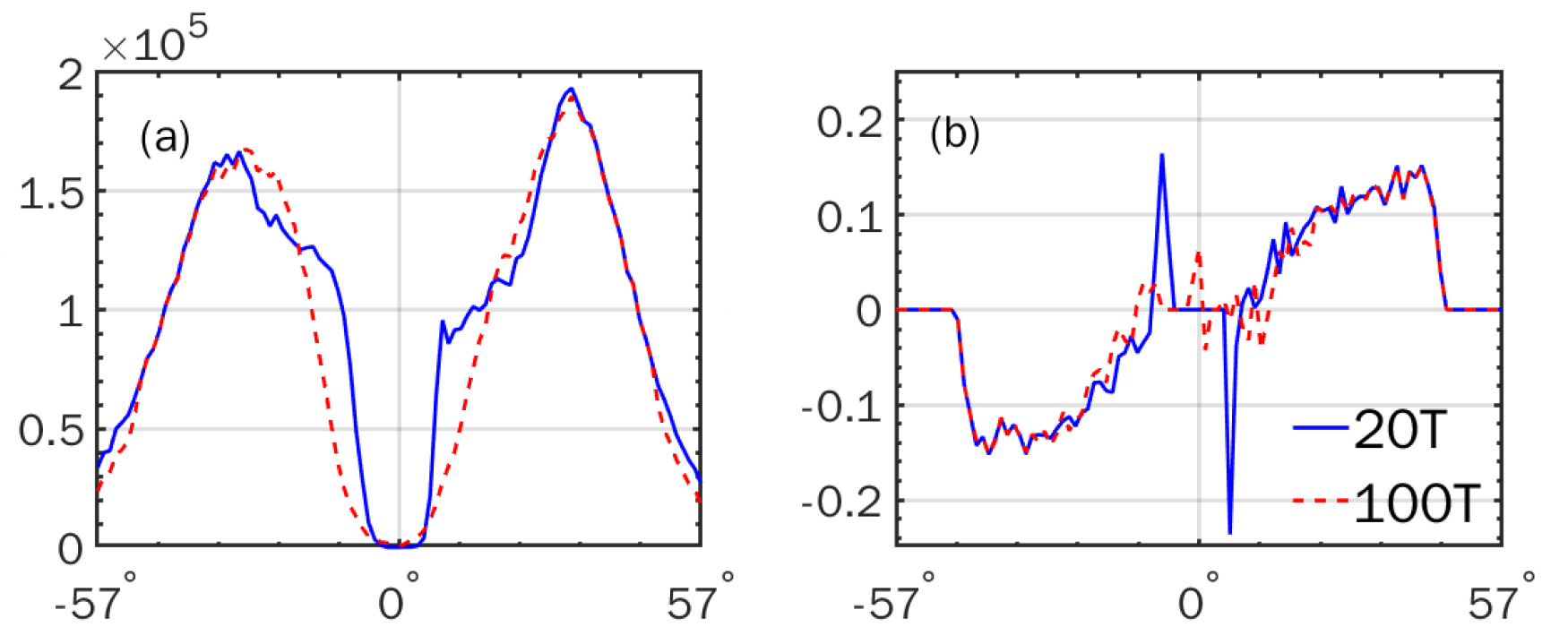}
\begin{picture}(300,-30)
   \put(62,3){{\small $\theta$}}
  \put(192,3){{\small $\theta$}}
  \put(0,53){\rotatebox{90}{\small $dN/d\theta$}}
  \put(130,67){\rotatebox{90}{\small $\zeta_y$}}
 \end{picture}
	\caption{The particle number distribution d$N$/d$\theta$ (a), and the averaged spin component $\zeta_y$ (b) at $t=20T$ (solid blue line) and $t=100T$ (dash red line).}
	\label{Fig.100T}
\end{figure}

\subsubsection{Small angle electrons: Angle-dependent polarization}

Now let us discuss the polarization of the electrons in a small angle region. At $t=20T$, the reflected electrons could either be scattered out of the laser beam in the transverse direction or still propagate forward along with the laser beam. The former corresponds to the case of the large-angle electrons described above, while the later to the small-angle electrons which have opposite polarization [see Fig. \ref{Fig. agl}].
To analyze polarization features  in the small angle region,
we select the electrons with final polar angle within $[-10^\circ, 10^\circ]$ and plot their trajectory, as well as the evolution of momentum, see Fig. \ref{Fig. small}. Apparently, the small-angle electrons are initially distributed around the beam center with $\overline{x}(0)\sim0$, see Fig.~\ref{Fig. small} (a) and (b). Unlike the off-axis electrons, the on-axis electrons are less likely to be scattered out of the beam when the laser peak arrives.
When the on-axis electrons meet the negative half-cycle peaked at $\varphi=4.55$, they are accelerated to relativistic velocity and stay in the half-cycle $4.55<\varphi<5.05$ for a few periods [see Fig.~\ref{Fig. small} (c) and (d)], during which the electrons move towards $x>0$ with $p_x>0$ [see Fig. \ref{Fig. small} (e) and (f)]. The radiative polarization still is non-negligible at this point as $\chi_e\sim0.45$, and induces polarization $\zeta_y>0$  correlated with momentum $p_x>0$. Unlike the large-angle electrons, the small-angle electrons are not scattered off the laser beam but continue to oscillate inside the laser beam. In the next half-cycle, the momentum $p_x$ becomes negative, but the polarization $\zeta_y$ couldn't keep up with the changes of $p_x$  as the radiative polarization is gradually suppressed along with decreasing $\chi_e$, see Fig. \ref{Fig. small} (g) and (h).
For instance, at the detection time $t=20T$,  the electrons with $p_x^f<0$ are centred at $\varphi=5.156$, they have $p_x<0$  [Fig. \ref{Fig. small} (f)] but still $\zeta_y>0$ [Fig. \ref{Fig. small} (h)]. Therefore, after reflection, the polarization of the small-angle electrons is barely changed as $\chi_e\rightarrow 0$ while the momentum $p_x$ further evolves with laser fields, which breaks the correlation between $p_x$ and $\zeta_y$. At $t=20T$, the dephasing between $p_x$ and $\zeta_y$ causes the opposite polarization for electrons distributed at small and large angles. However,  the angle-dependent polarization observed at $t=20T$ is not stable for small angle electrons, since $p_x$ further evolves with time as long as the electrons stay in the laser field. The electrons with different polarization could be mixed with each other, leading to a vanishing polarization for small angle electrons at  $t=100T$. In contrast, the electrons in a large angle region are free from depolarization as they are already out of laser pulse at  $t=20T$ [see Fig. \ref{Fig.100T}].

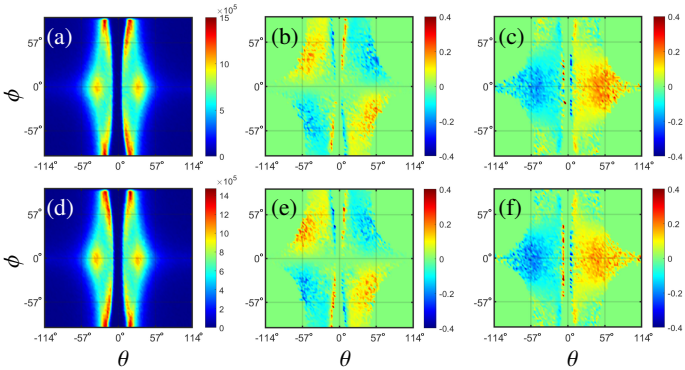
\begin{figure}[b]
	\includegraphics[width=0.5\textwidth]{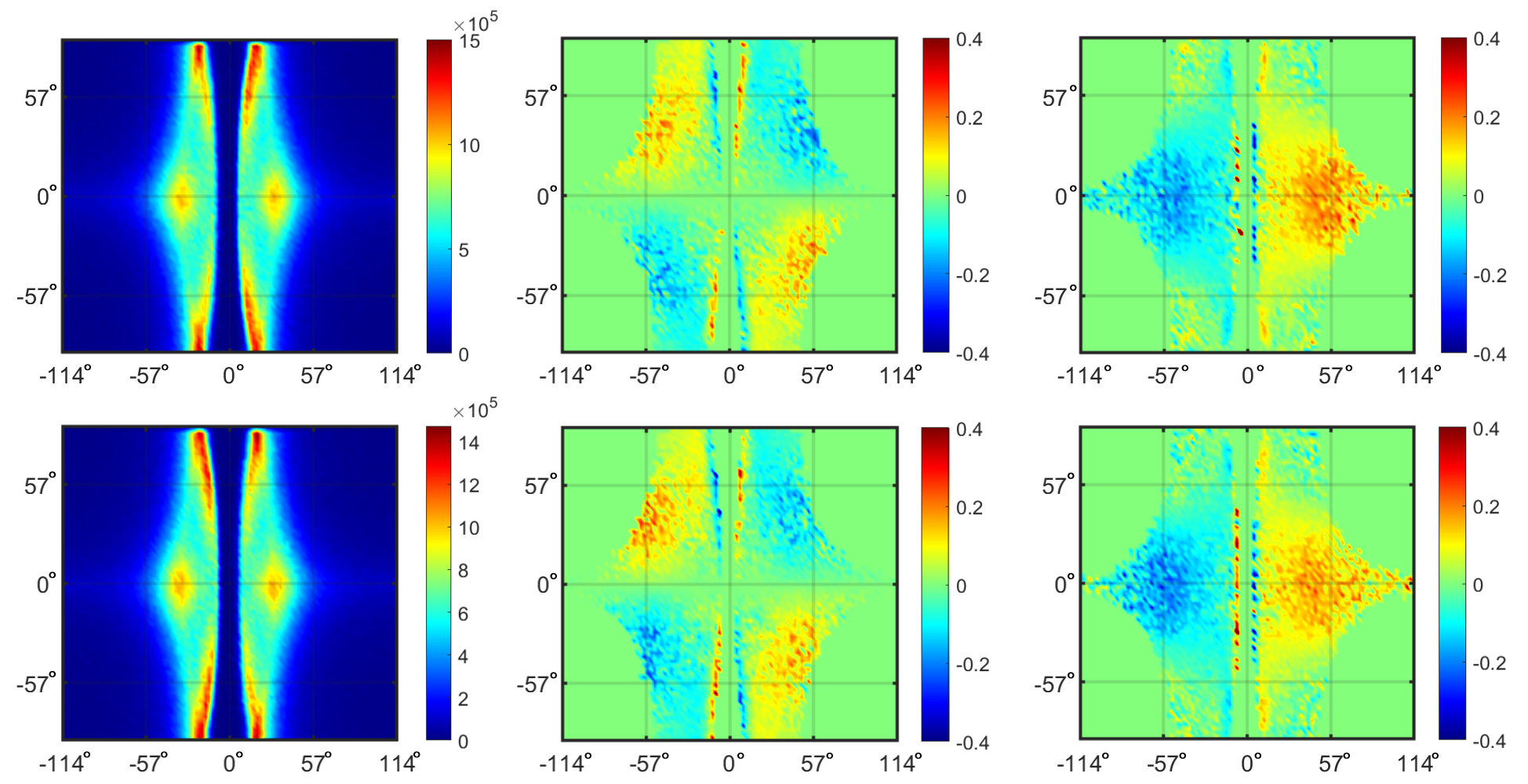}
\begin{picture}(300,-30)
\put(11,125){\small\textcolor{white}{(a)}}
  \put(96,125){\small(b)}
    \put(182,125){\small (c)}
  \put(11,60){\small\textcolor{white}{(d)}}
  \put(96,60){\small (e)}
  \put(182,60){\small(f)}
   \put(38,3){{\small $\theta$}}
   \put(122,3){{\small $\theta$}}
  \put(208,3){{\small $\theta$}}
  \put(-5,40){\rotatebox{90}{\small $\phi$}}
  \put(-5,103){\rotatebox{90}{\small $\phi$}}
 \end{picture}
	\caption{ Angular distribution  d$^2N$/d$\Omega$ vs the polar angle $\theta$ (degree) and the azimuthal angle $\phi$ (degree): (a) for produced electrons $e^-$ and (d) for positrons $e^+$. The averaged polarization distribution along the electric field direction $\zeta_x$: (b) for $e^-$  and  (e) for $e^+$. The averaged polarization distribution along the magnetic field direction $\zeta_y$: (c) for $e^-$  and (f) for $e^+$.}
	\label{Fig. pairs}
\end{figure}

\subsection{Polarization of created pairs}

The strong interaction induces efficient  pair production, the number of emitted pairs per laser cycle is \cite{Ritus_1985}:
\begin{eqnarray}
\label{Npp}
 N_{e+e-}\approx N_\gamma\frac{27\Gamma^7(2/3)\alpha m\lambda}{56\pi^5\/\omega_\gamma\lambda_C}\left(\frac{3\chi_\gamma}{2}\right)^{2/3}.
\end{eqnarray}
When each electron during a single laser cycle emits one high energy photon $N_\gamma\sim \alpha a_0\sim1$, for our parameters one can achieve $N_{e+e-}\sim 1$ per laser cycle via Eq.~(\ref{Npp}).
In qualitative terms this means that each electron emits one high-energy photon, which further is converted to an electron-positron pair in one laser cycle, yielding one positron per each initial electron. In this case, the positron number is comparable with the number of seed electrons (this yield is about two orders of magnitude larger than that of Ref.~\cite{chen2019polarized} using a two-color scheme).

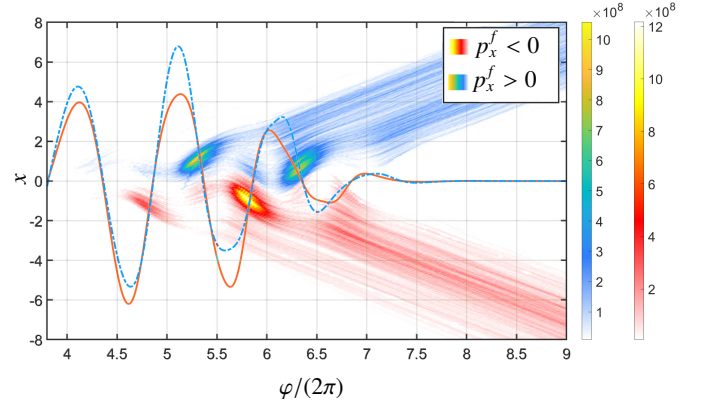
\begin{figure}[]
    \includegraphics[width=0.5\textwidth]{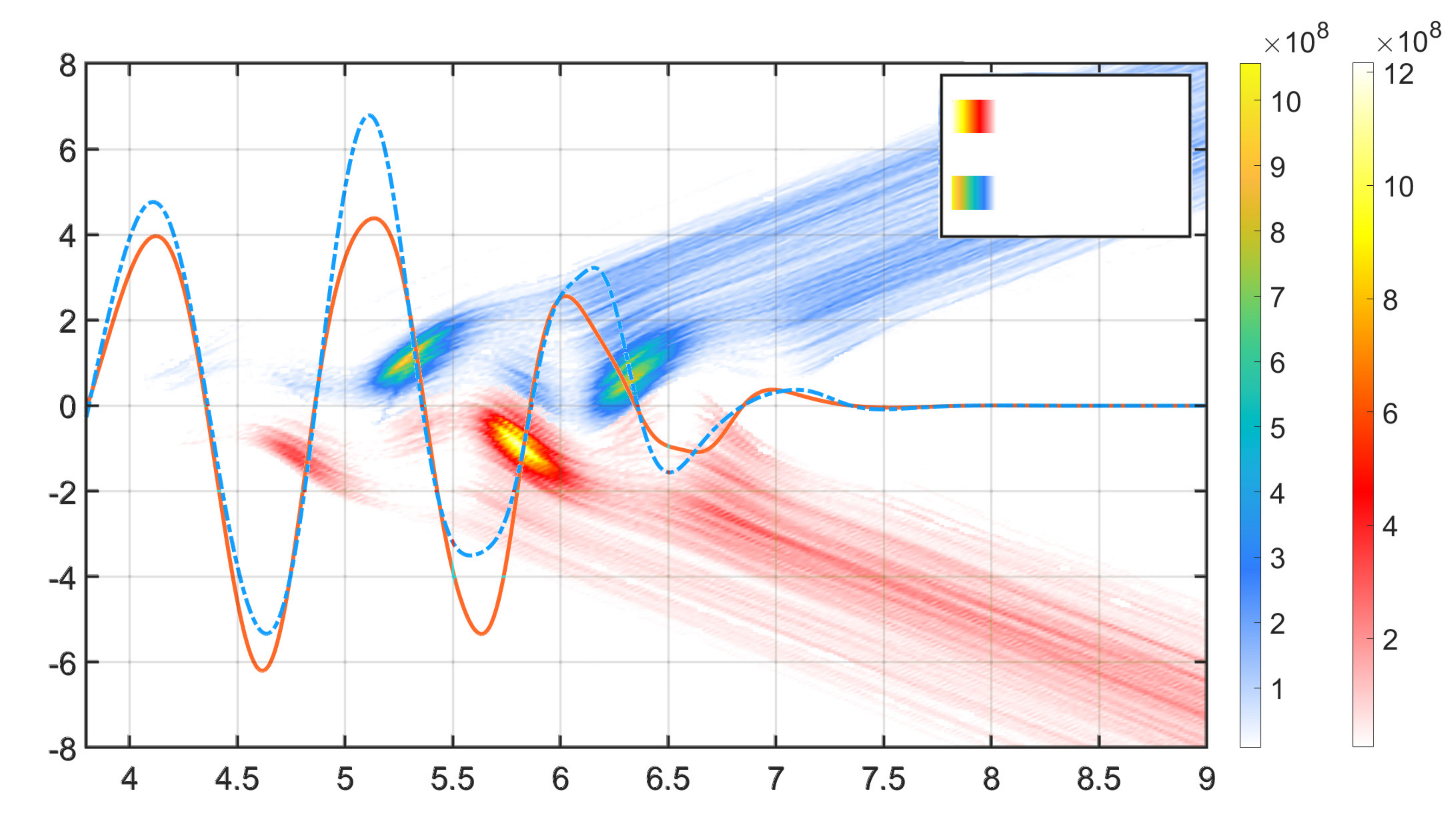}
\begin{picture}(300,-30)
  \put(103,3){{\small $\varphi/(2\pi)$}}
  \put(3,82){{\rotatebox{90}{$x$}}}
  \put(177,132){{\small $p_x^f<0$}}
  \put(176,119){{\small $p_x^f>0$}}
  \end{picture}
	\caption{The trajectory $x$ of positrons within $|\theta|\in[50^\circ,60^\circ]$ and $\phi\in[-20^\circ,20^\circ]$ vs  the laser phase $\varphi$ for positrons with $p_x^f>0$ (parula) and $p_x^f<0$ (hot). The superimposed lines are the corresponding magnetic fields.}
	\label{Fig.pair_tra}
\end{figure}

Similar to the seed electrons, the created pairs are polarized with $\zeta_y>0$ at $\theta>0$ and $\zeta_y<0$ at $\theta<0$, but with more particles distributed at $\theta>20^\circ$ [see Fig. \ref{Fig. pairs}]. This is because the created pairs have a larger $\overline{x}_\pm(0)$ than that of initial electrons, therefore more likely to be scattered out of the laser beam than captured by it.
The polarization mechanism for seed electrons also works for the produced pairs.
The created electrons and positrons are polarized according to the quantization axis in the creation phase. Positron (electron) polarization is most probably to be along (opposite to) the laser magnetic field.  Therefore, the positrons (electrons) created at $B_y>0$ are likely to be polarized with $\zeta_y^+>0$ ($\zeta_y^-<0$).
After creation, these positrons  (electrons)  with $\zeta_y^+>0$  ($\zeta_y^-<0$) lost most of their energy due to significant radiation at the half cycle $B_y>0$, since the radiation probability  $dW_{\zeta_i,\zeta_f}$  for positrons  (electrons) is dominated by $dW_{\uparrow,\uparrow}$ ($dW_{\downarrow,\downarrow}$). The reflected positrons (electrons) are rapidly accelerated to relativistic velocity and stay at the acceleration phase $B_y>0$, $p^+_x>0$ ($B_y>0$, $p^-_x<0$) for a long time, during which the positrons (electrons) with $x(0)^+>0$ ($x(0)^-<0$) can be rapidly scattered out of the laser beam [Fig. \ref{Fig.pair_tra}].
Consequently, the positrons (electrons) created at $B_y>0$, $\zeta_y^+>0$ ($\zeta_y^-<0$) have final momentum $p_x^f>0$ ($p_x^f<0$), while created at $B_y<0$, $\zeta_y^+<0$ ($\zeta_y^->0$) have $p_x^f<0$ ($p_x^f>0$).  Therefore, the angle distribution of density and polarization for pairs are similar to seed electrons, except for a higher density distribution in the large angle region.

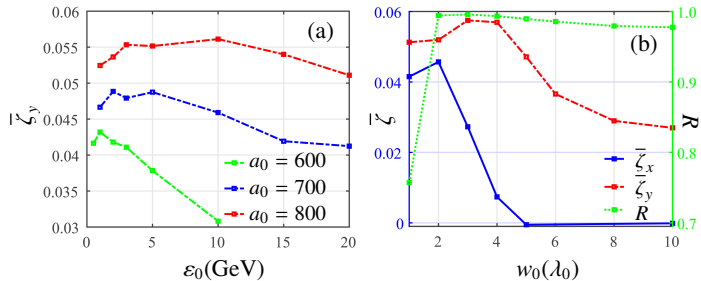
\begin{figure}[t]
    \includegraphics[width=0.5\textwidth]{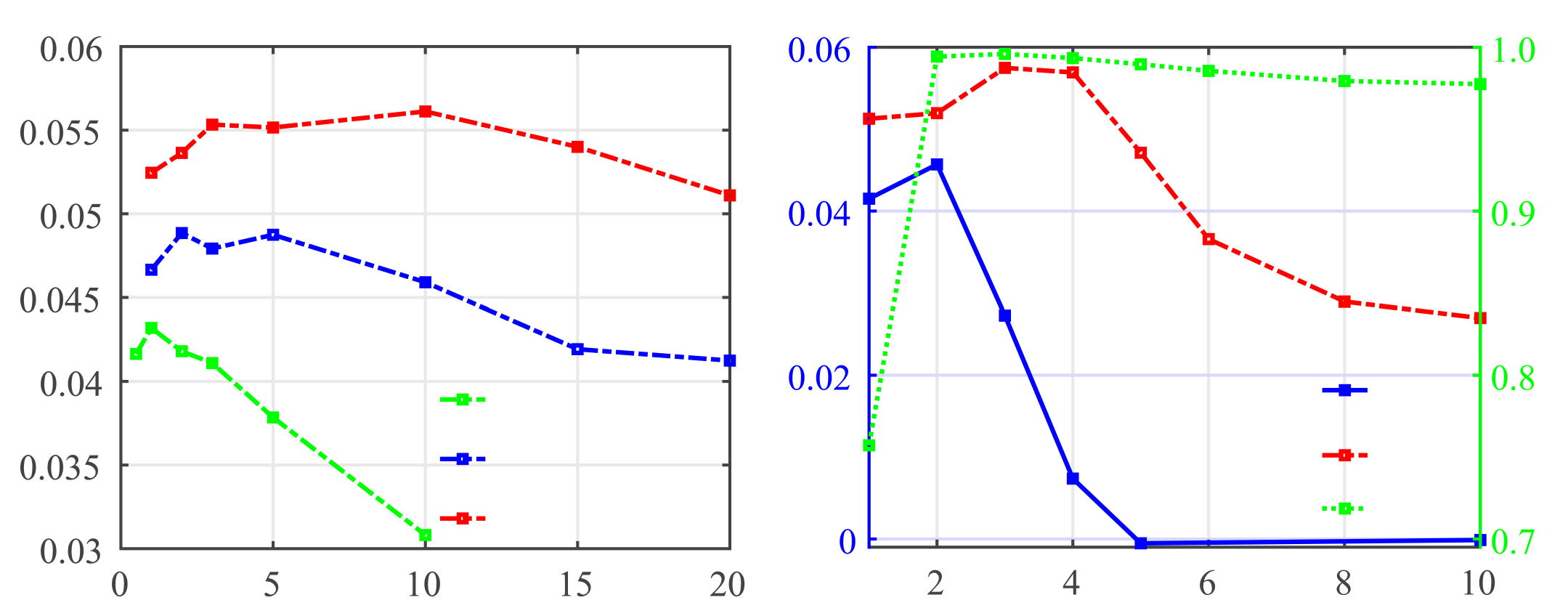}
\begin{picture}(300,-30)
  \put(103,93){\small (a)}
  \put(225,88){\small(b)}
  \put(-10,60){\rotatebox{90}{\small $\overline{\zeta}_y$}}
  \put(126,60){\rotatebox{90}{\small $\overline{\zeta}$}}
  \put(245,64){\rotatebox{-90}{\small $R$}}
  \put(56,3){{\small $\varepsilon_0$(GeV)}}
  \put(182,3){{\small $w_0(\lambda_0)$}}
  \put(81,44){{\footnotesize $a_0=600$}}
  \put(81,34){{\footnotesize $a_0=700$}}
  \put(81,24){{\footnotesize $a_0=800$}}
  \put(227,44){{\footnotesize $\overline{\zeta}_x$}}
  \put(227,34){{\footnotesize $\overline{\zeta}_y$}}
  \put(227,24){{\footnotesize $R$}}
  \end{picture}
	\caption{(a) The average polarization $\overline{\zeta}_y$ for electrons with $\theta>0$ vs seed electron energy $\varepsilon_0$ for $a_0=600$ (blue-solid), $a_0=700$ (red-dashed) and $a_0=800$ (green-dotted), respectively. (b) The average polarization $\overline{\zeta}_x$  at $\theta<0$, $\phi>0$ (blue-solid), $\overline{\zeta}_y$ at $\theta>0$ (red-dashed)  and  reflectivity  $R$ of the electron beam vs laser beam waist $w_0$, $a_0=760$.}
	\label{Fig.para}
\end{figure}

\section{Impact of laser and electron parameters}

For experimental convenience, we investigate the impact of laser and electron parameters on polarization, see Fig. \ref{Fig.para}. The average polarization $\overline{\zeta}_y$ is inversely proportional to the seed electrons energy [Fig. \ref{Fig.para} (a)]. This is because the energetic electrons can penetrate through the laser pulse instead of being reflected. Since the forward electrons have vanishing average polarization in a symmetric laser field, the average polarization decreases with higher electron energy. The dependence of  $\overline{\zeta}_y$  on $\varepsilon_0$ is less sensitive for larger $a_0$, since the radiation loss is more dramatic in an intense laser field which is beneficial for the occurrence of reflection and suppresses forward scattering.  Meanwhile, an intense laser pulse provides a higher transverse acceleration and stronger focusing effects, making it easier for electrons to escape the field with correlated polarization and deflection angle. Therefore, with the increase of $a_0$ the average polarization increases.  To show the importance of the focusing effect, we simulate the scattering with various laser beam waists $w_0$. Generally, the average polarization decreases with the increase of the beam waist due to the weaker focusing effects at larger $w_0$ [Fig. \ref{Fig.para} (b)].
 However, $\overline{\zeta}_y$ increases with the growth of $w_0$ at $w_0\leq 3\lambda_0$, which can be explained as follows. As is shown in Sec. \ref{sx}, the pondermotive force in a tightly focused laser beam  induces a rotation of the polarization vector. When the beam waist decreases from $3\lambda_0$ to $2\lambda_0$, the enhanced pondermotive force causes the decrease of $\overline{\zeta}_y$ and increase of $\overline{\zeta}_x$ [Fig. \ref{Fig.para} (b)].  Meanwhile, when the laser beam waist is further decreased, part of the electron beam may not be able to interact with strong lasers and consequently cannot be reflected. The reflection rate decreases to $76\%$ at $w_0=\lambda_0$, resulting in a decrease of $\overline{\zeta}_x$ and $\overline{\zeta}_y$ [Fig. \ref{Fig.para} (b)]. Therefore, $\overline{\zeta}_y$ increases with  $w_0$ at $w_0\leq 3\lambda_0$ as a result of enhanced spin precession and a reduction in the reflection rate.\\

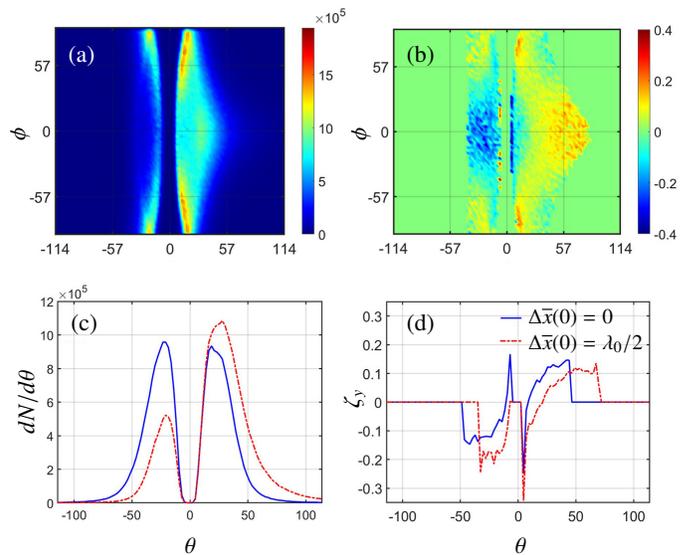
\begin{figure}[]
 \includegraphics[width=0.5\textwidth]{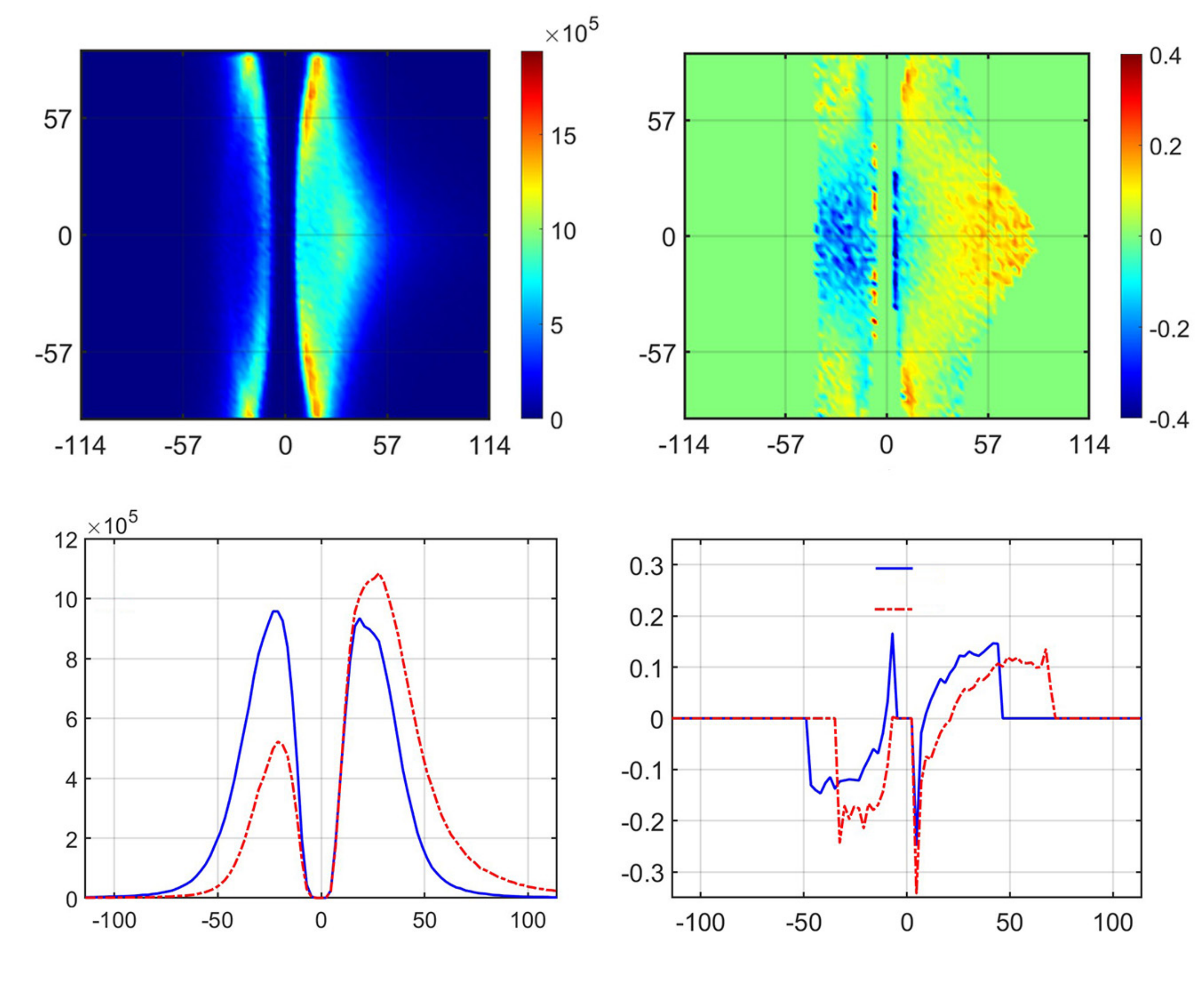}
\begin{picture}(300,-30)
  \put(22,196){\small\textcolor{white}{(a)}}
   \put(150,196){\small(b)}
   \put(150,94){\small(d)}
  \put(23,94){\small(c)}
   \put(196,97){{\footnotesize $\Delta \overline{x}(0)=0$}}
  \put(196,87){{\footnotesize $\Delta \overline{x}(0)=\lambda_0/2$}}
   \put(66,10){{\small $\theta$}}
  \put(191,10){{\small $\theta$}}
  \put(0,166){\rotatebox{90}{\small $\phi$}}
  \put(129,166){\rotatebox{90}{\small $\phi$}}
  \put(3,55){\rotatebox{90}{\small $dN/d\theta$}}
  \put(126,65){\rotatebox{90}{\small $\zeta_y$}}
 \end{picture}
 \caption{Angular distribution of $d^2N/d\Omega$ (a) and averaged polarization distribution along the magnetic field direction $\zeta_y$ (b) versus the polar angle $\theta$ (degree) and the azimuthal angle $\phi$ (degree) for electrons with an initial displacement of $\Delta \overline{x}(0)=\lambda_0/2$. The electron number distribution $dN/d\theta$ (c) and the averaged spin component $\zeta_y$ (d) versus polar angle $\theta$  for electron beam with (solid blue line) and without (dash red line) initial beam displacement.}
 \label{Fig. shift}
\end{figure}

{\color{black}The impact of initial transverse displacement of the electron beam  is analyzed in Fig. \ref{Fig. shift}.  For $\Delta \overline{x} (0)=\lambda_0/2$, the symmetric angular distribution of electrons is distorted, with more electrons moving towards $\theta>0$ and $\zeta_y>0$, see Fig. \ref{Fig. shift}. The electrons with larger initial displacement are more likely to be scattered out of laser field, and consequently more electrons are distributed in the larger angle region [Fig. \ref{Fig. shift} (a) and (c)], where the polarization degree is higher [Fig. \ref{Fig. shift} (b) and (d)].
Therefore,  a finite displacement of the electron beam from the laser pulse axis is beneficial for polarization. However, $\Delta \overline{x}(0)$ should be restricted, because the electrons far away from the laser beam axis would experience a rather weak field, which will reduce the average polarization of the beam.}

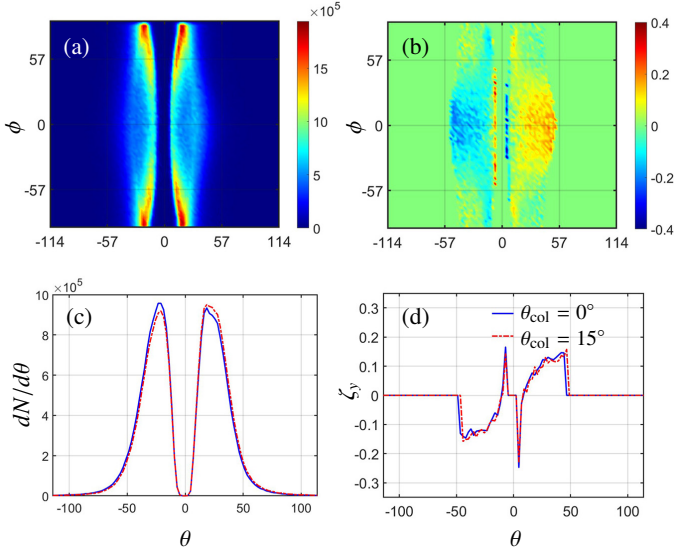
\begin{figure}[]
 \includegraphics[width=0.5\textwidth]{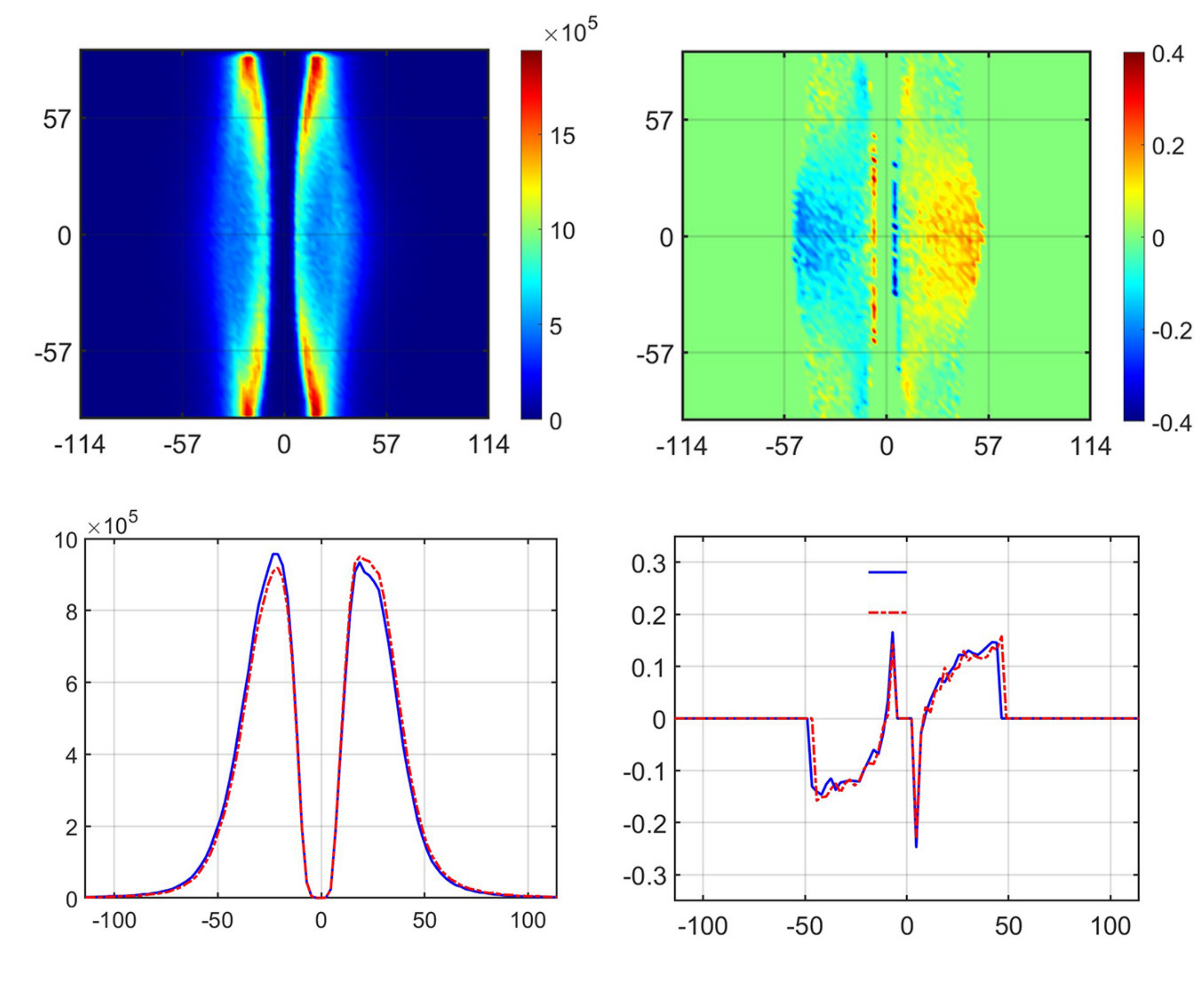}
\begin{picture}(300,-30)
  \put(22,196){\small\textcolor{white}{(a)}}
   \put(150,196){\small(b)}
   \put(150,94){\small(d)}
  \put(23,94){\small(c)}
  \put(195,96){{\footnotesize $\theta_{\text{col}}=0^\circ$}}
  \put(195,86){{\footnotesize $\theta_{\text{col}}=15^\circ$}}
   \put(66,10){{\small $\theta$}}
  \put(191,10){{\small $\theta$}}
  \put(0,166){\rotatebox{90}{\small $\phi$}}
  \put(129,166){\rotatebox{90}{\small $\phi$}}
  \put(3,55){\rotatebox{90}{\small $dN/d\theta$}}
  \put(126,65){\rotatebox{90}{\small $\zeta_y$}}
 \end{picture}
 \caption{Angular distribution of $d^2N/d\Omega$ (a) and averaged polarization distribution along the magnetic field direction $\zeta_y$ (b) versus the polar angle $\theta$ (degree) and the azimuthal angle $\phi$ (degree) for electrons with a collision angle $\theta_{\text{col}}=15^\circ$ and displacement $\Delta \overline{x}(0)=-\lambda_0/2$. The electron number distribution $dN/d\theta$ (c) and the averaged spin component $\zeta_y$ (d) versus polar angle $\theta$  for electron beam with  $\theta_{\text{col}}=0^\circ$ (solid blue line) and  $\theta_{\text{col}}=15^\circ$  (dash red line).}
 \label{Fig. collision}
\end{figure}

{\color{black}Finally, we have investigated the impact of the collision angle between laser pulse and electron beam axis, see Fig. \ref{Fig. collision}. We rotate the electron beam  clockwise in the $x-z$ plane around the beam center with $\theta_{\text{col}}=15^\circ$, and shift the beam downwards with a displacement of $\Delta \overline{x}(0)=-\lambda_0/2$ for a collision. As discussed above, the negative displacement causes more electrons to be deflected towards $\theta<0$ and $\zeta_y<0$. However, some particles starting with negative displacement along the $x$ axis can enter the region of positive displacement due to the initial momentum. The initial momentum cancels out the polarization effects of displacement, leading to a negligible variation of angular distribution [Fig. \ref{Fig. collision} (c) and (d)]. Therefore, our results are robust with regard to the collision angle.}

\section{Conclusion}\label{concl}

We have demonstrated that an electron beam counterpropagating an ultrastrong ultrashort laser pulse can be polarized in the QRDR regime due to reflections. With a similar mechanism the created electron-positron pairs during the interaction are polarized, resulting in the generation of a dense beam of polarized particles with a linearly polarized laser pulse.
Strong spin-dependent radiation reaction is essential to enable the reflection scenario and creates angle-dependent polarization in tightly focused laser fields. The tightly  focusing of the laser beam and its ultrashort duration are necessary ingredients to enable the polarization mechanism. After the interaction of reflected particles with the dominant half-cycle, the field intensity is dramatically decreased due to the focusing effect of the laser field, resulting in the asymmetric and correlated kinetic and spin dynamics. Particles distributed with large transverse position, can be scattered out of the laser beam rapidly within the acceleration cycle, leading to the angle-dependent polarization along the laser magnetic field.

Furthermore, we have found that the spin precession also plays a role for developing  polarization features of the reflected particles. It reduces the polarization degree  along the magnetic field induced by the radiative polarization, diverting it to the angle-dependent polarization along the electric field.
In the same setup efficient pair production and their polarization with the same mechanism takes place.
For instance, with a possible ultrahigh-charge ($\sim$100 nC)  electron beam,  a polarized positron beam with a density of $10^9-10^{10}$/bunch is foreseeable.

The important point of the presented reflection scheme of the particle polarization is that the polarization arises from the asymmetric laser-electron interaction dynamics, and it does not demand the challenging task of constructing asymmetric strong laser fields.
With next generation laser facilities, the reflection scheme may provide a relatively simple way of producing dense polarized lepton beams and a valuable tool to search for qualitative signatures of radiative spin effects.\\

 \section*{Acknowledgement}  This work has been supported by the National Natural Science Foundation of China (Grant No. 12074262 and No. 12222507), the National Key R$\&$D Program of China (2021YFA1601700), the Shanghai Rising-Star Program, and the Shanghai Natural Science Foundation (Grant No. 20ZR1441600).

{\color{black}
\section*{Appendix A: spin- and polarization-resolved Monte-Carlo method for nonlinear Compton scattering}
In this section, we summarize the spin- and polarization-resolved Monte-Carlo method for nonlinear Compton scattering, which was developed in our previous works \cite{chen2022electron,dai2022photon,li2020polarized,li2022helicity,li2020production}
(also see \cite{song2022dense} for a similar algorithm).
\subsection*{A1: spin- and polarization-resolved photon emission probability}
The radiation probability including all the polarization and spin characteristics takes the form
\begin{align}\label{PRB_tot_angle}
 dW^R\left(\bm{\zeta},\bm{\zeta}',\bm{\xi}\right)=\frac{1}{2}\left(F_0+\xi_1F_1+\xi_2F_2+\xi_3F_3\right),
\end{align}
where the 3-vector $\bm{\xi}=\left(\xi_1,\xi_2,\xi_3\right)$ is the Stokes parameter of the emitted photon, $F_0= dW_{11}+ dW_{22}$, $F_1= dW_{12}+ dW_{21}$, $F_2=i\left( dW_{12}- dW_{21}\right)$, $F_3= dW_{11}- dW_{22}$, with the  polarization matrix of radiation probability per unit time:
\begin{align*}
dW_{11}+ dW_{22} & =\frac{C_0}{2}d\omega\left\{ \left[\frac{\varepsilon^{2}+\varepsilon'^{2}}{\varepsilon'\varepsilon}\textrm{K}_{\frac{2}{3}}\left(z_{q}\right)-\int_{z_q}^{\infty}dx\textrm{K}_{\frac{1}{3}}\left(x\right)\right]\right.\\
 & +\left[2\textrm{K}_{\frac{2}{3}}\left(z_{q}\right)-\int_{z_{q}}^{\infty}dx\textrm{K}_{\frac{1}{3}}\left(x\right)\right]\bm{\zeta}\cdot\bm{\zeta}'\\
 & -\left[\frac{\omega}{\varepsilon}\bm{\zeta}\cdot\mathbf{b}+\frac{\omega}{\varepsilon'}\bm{\zeta}'\cdot\mathbf{b}\right]\textrm{K}_{\frac{1}{3}}\left(z_{q}\right)\\
 & +\left.\frac{\omega^2}{\varepsilon'\varepsilon}\left[\textrm{K}_{\frac{2}{3}}\left(z_{q}\right)-\int_{z_{q}}^{\infty}dx\textrm{K}_{\frac{1}{3}}\left(x\right)\right]\left(\bm{\zeta}\cdot\mathbf{\hat{v}}\right)\left(\bm{\zeta}'\cdot\mathbf{\hat{v}}\right)\right\},
\end{align*}
\begin{align*}
dW_{12}+dW_{21}&=\frac{C_{0}}{2}d\omega\left\{ \frac{\varepsilon^{2}-\varepsilon'^{2}}{2\varepsilon'\varepsilon}\textrm{K}_{\frac{2}{3}}\left(z_{q}\right)\left(\hat{\mathbf{v}}\left[\bm{\zeta}'\times\bm{\zeta}\right]\right)\right.\\&+\left[\frac{\omega}{\varepsilon'}\left(\bm{\zeta}\cdot\mathbf{\mathbf{s}}\right)+\frac{\omega}{\varepsilon}\left(\bm{\zeta}'\cdot\mathbf{s}\right)\right]\textrm{K}_{\frac{1}{3}}\left(z_{q}\right)\\&-\frac{\omega^{2}}{2\varepsilon'\varepsilon}\int_{z_{q}}^{\infty}dx\textrm{K}_{\frac{1}{3}}\left(x\right)\left[\left(\bm{\zeta}\cdot\mathbf{s}\right)\left(\bm{\zeta}'\cdot\mathbf{b}\right)\right.\\&+\left.\left.\left(\bm{\zeta}\cdot\mathbf{b}\right)\left(\bm{\zeta}'\cdot\mathbf{s}\right)\right]\right\},
\end{align*}
\begin{align*}
dW_{12}-dW_{21} & =i\frac{C_0}{2}d\omega\left\{ \frac{\varepsilon^{2}-\varepsilon'^{2}}{2\varepsilon'\varepsilon}\textrm{K}_{\frac{1}{3}}\left(z_{q}\right)\left(\mathbf{s}\cdot\left[\bm{\zeta}'\times\bm{\zeta}\right]\right)\right.\\
 & +\left(-\frac{\varepsilon^{2}-\varepsilon'^{2}}{\varepsilon'\varepsilon}\textrm{K}_{\frac{2}{3}}\left(z_{q}\right)+\frac{\omega}{\varepsilon}\int_{z_{q}}^{\infty}dx\textrm{K}_{\frac{1}{3}}\left(x\right)\right)\left(\bm{\zeta}\cdot\mathbf{\hat{v}}\right)\\
 & +\left(-\frac{\varepsilon^{2}-\varepsilon'^{2}}{\varepsilon'\varepsilon}\textrm{K}_{\frac{2}{3}}\left(z_{q}\right)+\frac{\omega}{\varepsilon'}\int_{z_{q}}^{\infty}dx\textrm{K}_{\frac{1}{3}}\left(x\right)\right)\left(\bm{\zeta}'\cdot\mathbf{\hat{v}}\right)\\
 & +\left.\frac{\omega^{2}}{2\varepsilon'\varepsilon}\textrm{K}_{\frac{1}{3}}\left(z_{q}\right)\left[\left(\bm{\zeta}\cdot\mathbf{\hat{v}}\right)\left(\bm{\zeta}'\cdot\mathbf{b}\right)+\left(\bm{\zeta}\cdot\mathbf{b}\right)\left(\bm{\zeta}'\cdot\mathbf{\hat{v}}\right)\right]\right\},
\end{align*}
\begin{align}\nonumber \label{PRB_ELE}
dW_{11}-dW_{22}&=\frac{C_{0}}{2}d\omega\left\{ \textrm{K}_{\frac{2}{3}}\left(z_{q}\right)+\frac{\varepsilon^{2}+\varepsilon'^{2}}{2\varepsilon'\varepsilon}\textrm{K}_{\frac{2}{3}}\left(z_{q}\right)\bm{\zeta}\cdot\bm{\zeta}'\right.\\\nonumber&-\left[\frac{\omega}{\varepsilon'}\left(\bm{\zeta}\cdot\mathbf{b}\right)+\frac{\omega}{\varepsilon}\left(\bm{\zeta}'\cdot\mathbf{b}\right)\right]\textrm{K}_{\frac{1}{3}}\left(z_{q}\right)\\\nonumber&+\frac{\omega^{2}}{2\varepsilon'\varepsilon}\left(-\textrm{K}_{\frac{2}{3}}\left(z_{q}\right)\left(\bm{\zeta}\cdot\mathbf{\hat{v}}\right)\left(\bm{\zeta}'\cdot\mathbf{\hat{v}}\right)\right.\\\nonumber&+\int_{z_{q}}^{\infty}dx\textrm{K}_{\frac{1}{3}}\left(x\right)\left[\left(\bm{\zeta}\cdot\mathbf{b}\right)\left(\bm{\zeta}'\cdot\mathbf{b}\right)\right.\\&-\left.\left.\left.\left(\bm{\zeta}\cdot\mathbf{\mathbf{s}}\right)\left(\bm{\zeta}'\cdot\mathbf{s}\right)\right]\right)\right\} ,
\end{align}
where $\bm{\zeta}$ and  $\bm{\zeta}'$ are the spin polarization vectors before and after emission, $\varepsilon$ and $\varepsilon'$ the corresponding electron energy, $z_q=\frac{2}{3\chi_e}\frac{\omega}{\varepsilon'}$ with $\chi_e$ controlling the magnitude of quantum effects, $C_0=\frac{\alpha}{\sqrt{3}\pi\gamma^2}$ and $\mathbf{b}=\mathbf{\hat{v}}\times \bm{s}$ with $\mathbf{\hat{v}}$ and $\bm{s}$ being unit vectors along the direction of electron velocity and acceleration, respectively. \\

\noindent
\textit{Spin quantization axis for final electron spin}: After summing over the  polarization of emitted photons, we obtain the probability for  emitting a photon with an energy $\omega$ and spin $\bm{\zeta}'$ during the time step $\Delta t$:
\begin{align}\nonumber
&d\overline{W}^R(\bm{\zeta},\bm{\zeta}')  =\frac{1}{2} \left(a+\bm{b}\cdot\bm{\zeta}'\right)\\\nonumber
&a =C_0 d\omega\left\{ \frac{\varepsilon^{2}+\varepsilon'^{2}}{\varepsilon'\varepsilon}\textrm{K}_{\frac{2}{3}}\left(z_{q}\right)-\intop_{z_q}^{\infty}dx\textrm{K}_{\frac{1}{3}}\left(x\right)-\frac{\omega}{\varepsilon}\bm{\zeta}\mathbf{b}\textrm{K}_{\frac{1}{3}}\left(z_{q}\right)\right\},\\\nonumber
 &\bm{b}=C_0 d\omega\left\{
 \left[2\textrm{K}_{\frac{2}{3}}\left(z_{q}\right)-\intop_{z_{q}}^{\infty}dx\textrm{K}_{\frac{1}{3}}\left(x\right)\right]\bm{\zeta}-\frac{\omega}{\varepsilon'}\textrm{K}_{\frac{1}{3}}\left(z_{q}\right)\mathbf{b}\right.\\
 & +\left.\frac{\omega^2}{\varepsilon'\varepsilon}\left[\textrm{K}_{\frac{2}{3}}\left(z_{q}\right)-\intop_{z_{q}}^{\infty}dx\textrm{K}_{\frac{1}{3}}\left(x\right)\right]\left(\bm{\zeta}\mathbf{\hat{v}}\right)\mathbf{\hat{v}}\right\}.
\end{align}
The final
polarization vector of the electron resulting from the scattering process itself is
$\bm{\zeta}_f^R=\frac{\bm{b}}{a}$, which determines the spin quantization axis for electrons after radiation adopted in our Monte-Carlo simulations: $\bm{n}^R=\bm{\zeta}_f^R/|\bm{\zeta}_f^R|$.\\

\noindent
\textit{Polarization quantization axis for emitted photon}: Summing over the final electron polarizations, the radiation probability becomes
\begin{equation}\nonumber
d\widetilde{W}^R\left(\bm{\zeta},\bm{\xi}\right)=\frac{1}{2}\left(\widetilde{F}_0+\xi_1\widetilde{F}_1+\xi_2\widetilde{F}_2+\xi_3\widetilde{F}_3\right),
\end{equation}
\begin{align*}
\widetilde{F}_0 & =C_0d\omega\left\{ \frac{\varepsilon^{2}+\varepsilon'^{2}}{\varepsilon'\varepsilon}\textrm{K}_{\frac{2}{3}}\left(z_{q}\right)-\int_{z_q}^{\infty}dx\textrm{K}_{\frac{1}{3}}\left(x\right)-\frac{\omega}{\varepsilon}\bm{\zeta}\cdot\mathbf{b}\textrm{K}_{\frac{1}{3}}\left(z_{q}\right)\right\},
\end{align*}
\begin{align*}
\widetilde{F}_1 & =C_0d\omega \frac{\omega}{\varepsilon'}\left(\bm{\zeta}\cdot\mathbf{\mathbf{s}}\right)\textrm{K}_{\frac{1}{3}}\left(z_{q}\right),
\end{align*}
\begin{align*}
\widetilde{F}_2 & =-C_0d\omega
 \left(-\frac{\varepsilon^{2}-\varepsilon'^{2}}{\varepsilon'\varepsilon}\textrm{K}_{\frac{2}{3}}\left(z_{q}\right)+\frac{\omega}{\varepsilon}\int_{z_{q}}^{\infty}dx\textrm{K}_{\frac{1}{3}}\left(x\right)\right)\left(\bm{\zeta}\cdot\mathbf{\hat{v}}\right),
\end{align*}
\begin{align} \label{PRB_ELE2}
\widetilde{F}_3& =C_0d\omega\left\{ \textrm{K}_{\frac{2}{3}}\left(z_{q}\right)-\frac{\omega}{\varepsilon'}\left(\bm{\zeta}\cdot\mathbf{b}\right)\textrm{K}_{\frac{1}{3}}\left(z_{q}\right)\right\}.
\end{align}
The polarization of the emitted photon resulting from the scattering process itself takes the form $\bm{\xi}_f=\left(\widetilde{F}_1/\widetilde{F}_0,\widetilde{F}_2/\widetilde{F}_0,\widetilde{F}_3/\widetilde{F}_0\right)$, which determines the quantization axis for the emitted photons polarization: $\bm{n}_\gamma^R=\bm{\xi}_f/|\bm{\xi}_f|$.

After summing over final polarizations, we obtain the spectral probability depending on the initial spin of the electron $\bm{\zeta}$:
\begin{align}\label{WR0}
dW_T^R&\left(\bm{\zeta}\right)=\widetilde{F}_0.
\end{align}

\subsection*{A2: Spin-resolved no-emission probability}
While electron polarization emerges mostly due to spin-flips at photon emissions, there is a nonradiative contribution to the polarization which stems from the one-loop QED radiative corrections to the self-energy. In physical terms, the nonradiative polarization effect emerges due to the dependence of the photon emission probability on the initial electron spin. 
The electrons which do not emit will be polarized, because the emission is preferred in a certain spin state.
For a detailed discussions on this topic, please refer to our recent paper \cite{li2022strong}.

The spin-resolved no-emission probability can be derived from the emission probability Eq. (\ref{WR0}) based on unitarity:
\begin{align}\nonumber \label{PRB_NR}
&W^{NR}(\bm{\zeta},\bm{\zeta}')=\frac{1}{2}\left(c+\bm{\zeta}'\cdot{\bm d}\right),\\\nonumber
&c=1-\int_0^\varepsilon\widetilde{F}_0d\omega\Delta t,\\
&\bm{d}=\bm{\zeta}\left(1-\int_0^\varepsilon\overline{F}_0d\omega\Delta t\right)+\mathbf{b}C_0\int_0^\varepsilon\frac{\omega}{\varepsilon}\textrm{K}_{\frac{1}{3}}\left(z_{q}\right)d\omega\Delta t,
\end{align}
where
\begin{align*}
\overline{F}_0 & =C_0d\omega\left\{ \frac{\varepsilon^{2}+\varepsilon'^{2}}{\varepsilon'\varepsilon}\textrm{K}_{\frac{2}{3}}\left(z_{q}\right)-\intop_{z_q}^{\infty}dx\textrm{K}_{\frac{1}{3}}\left(x\right)\right\}.
\end{align*}
The final
polarization vector of the electron resulting from the scattering process itself is
$\bm{\zeta}_f^{NR}=\frac{\bm{d}}{c}$, which determines the spin quantization axis for electrons after no-emission process: $\bm{n}^{NR}=\bm{\zeta}_f^{NR}/|\bm{\zeta}_f^{NR}|$.

\subsection*{A3: Classical spin precession in the external laser field}

Between quantum events, the electron dynamics in the ultraintense laser field is described by the Lorenz equation
\begin{eqnarray}\label{L}
{\rm d}{\bf p}/{\rm d}t&=&e({\bf E+{\bm \beta}\times{\bf B}}).
\end{eqnarray}

The spin procession is governed by the Thomas-Bargmann-Michel-Telegdi equation:
\begin{eqnarray}\label{BMT}
\frac{{\rm d}{\bf S}}{{\rm d}t}&=&\frac{e}{m}{\bf S}\times\left[-\left(\frac{g}{2}-1\right)\frac{\gamma}{\gamma+1}\left({\bm \beta}\cdot{\bf B}\right){\bm \beta}\right.\nonumber\\
&&\left.+\left(\frac{g}{2}-1+\frac{1}{\gamma}\right){\bf B}-\left(\frac{g}{2}-\frac{\gamma}{\gamma+1}\right){\bm \beta}\times{\bf E}\right],
\end{eqnarray}
where ${\bf E}$ and  ${\bf B}$ are the  laser electric and magnetic fields, respectively, and
$g$ is the electron gyromagnetic factor:
$g\left(\chi_e\right)=2+2\mu\left(\chi_e\right)$, $\mu\left(\chi_e\right)=\frac{\alpha}{\pi\chi_e}\int_{0}^{\infty}\frac{y}{\left(1+y\right)^3}{\bf L}_{\frac{1}{3}} \left(\frac{2y}{3\chi_e}\right){\rm d}y$, with ${\bf L}_{\frac{1}{3}} \left(z\right)=\int_{0}^{\infty}{\rm sin}\left[\frac{3z}{2}\left(x+\frac{x^3}{3}\right)\right]{\rm d}x$. As $\chi_e\ll1$, $g\approx2.00232$.\\

\subsection*{A4: Algorithm of event generation}
1. \textit{Decide photon emission event}: At each simulation step, the photon emission and the photon energy are determined by the spectral probability  $dW^R_T$ of Eq. (\ref{WR0}), using the common stochastic procedure.

(1) Generate two random numbers $r_1,r_2\in [0,1]$ with uniform probability.

(2) Compute the radiation probability $P_m(r_1)$  for the given initial spin $\bm{\zeta}$ and photon energy  $\omega=r_1^3\varepsilon$.  The probability $P_m(r_1)$ in a given time interval $\Delta t$ is computed using  \cite{gonoskov2015extended}
\begin{equation}
P_m(r_1)=\frac{\partial f\left(r_{1}\right)}{\partial r_{1}}P\left[f\left(r_{1}\right)\right],
\end{equation}
with $f(x)=x^3$ and $P(\omega)=dW_T^R(\bm{\zeta},\omega)\Delta t$, i.e. $P(r_1)=3r_1^2dW_T^R(\bm{\zeta}, r_1^3\varepsilon)\Delta t$ with $\Delta t=10^{-3} T$.

(3) If $r_2<P_m(r_1)$, a photon is emitted with energy $\omega=r_1^3\varepsilon$. Otherwise, a photon emission is rejected.\\

2. \textit{Decide the polarization of outgoing particles}: \\

\textbf{Case 1}: $P_m(r_1)>r_2$: photon emission occurs. After each photon emission, the spin of the emitting particle (polarization of emitted photon) is either parallel or antiparallel to $\bm{n}^R$ ($\bm{n}_\gamma$) using the stochastic procedure with another random number $r_3\in[0,1]$.
For the given photon energy $\omega$ and initial spin $\bm{\zeta}$, compute the radiation probability $P_{\bm{\zeta'}\bm{\xi}}=dW^R(\bm{\zeta},\bm{\zeta}',\bm{\xi})\Delta t$. Here $\{\bm{\zeta}',\bm{\xi}\}\in \{\uparrow,\downarrow\}$ indicates the final spin is parallel or antiparallel with respect to the quantization axis.

(1) If $r_3<P_{\downarrow\downarrow}$, the electron is spin down with respect to $\bm{n}^R$ and the emitted photon is in the polarization state of $-\bm{n}_\gamma$.

(2) If $P_{\downarrow\downarrow}<r_3<P_{\downarrow\downarrow}+P_{\downarrow\uparrow}$, $\bm{\zeta}'=-\bm{n}^R$ and $\bm{\xi}=\bm{n}_\gamma$.

(3) If $P_{\downarrow\downarrow}+P_{\downarrow\uparrow}<r_3<P_{\downarrow\downarrow}+P_{\downarrow\uparrow}+P_{\uparrow\downarrow}$, $\bm{\zeta}'=\bm{n}^R$ and $\bm{\xi}=-\bm{n}_\gamma$.

(4) If $P_{\downarrow\downarrow}+P_{\downarrow\uparrow}+P_{\uparrow\downarrow}<r_3<P_{\downarrow\downarrow}+P_{\downarrow\uparrow}+P_{\uparrow\downarrow}+P_{\uparrow\uparrow}$, $\bm{\zeta}'=\bm{n}^R$ and $\bm{\xi}=\bm{n}_\gamma$.

\textbf{Case 2}: $P_m(r_1)<r_2$: photon emission is rejected. The electron spin state is collapsed into one of its basis states defined with respect to the instantaneous spin quantization axis $\bm{n}^{NR}$.

(1) Generate another random number $r_4\in[0,1]$.

(2) Compute $P_{\bm{\zeta}'}=W^{NR}\left(\bm{\zeta},\bm{\zeta}'\right)$ with $\bm{\zeta}'\in\{\uparrow,\downarrow\}$ indicating spin parallel or antiparallel with $\bm{n}^{NR}$.

(3) If $P_\uparrow/\left(P_\uparrow+P_\downarrow\right)>r_4$, $\bm{\zeta}'=\bm{n}^{NR}$; otherwise, $\bm{\zeta}'=-\bm{n}^{NR}$.\\

Note that, in our algorithm, the spin of the electron after the emission is determined by the spin-resolved emission probabilities according to the stochastic algorithm and instantaneously collapses into one of its basis states defined with respect to the instantaneous spin quantization axis (SQA). The pure state is more physically meaningful and more suitable to show a realistic detection result. Alternatively, one could set the final electron after emission in a mixed spin state $\bm{\zeta}'=\bm{\zeta}^R_f$ and photon polarization $\bm{\xi}=\bm{\xi}_f$; or $\bm{\zeta}'=\bm{\zeta}^{NR}_f$ in the case of no-emission.
The mixed state is more relevant for describing the average polarization of electron ensembles or macro-particles in PIC. Nevertheless, these two methods are equivalent for sufficient running \cite{tang2021radiative}.\\

\section*{Appendix B: Spin- and polarization-resolved Monte-Carlo Method for Breit-Wheeler process}
In this section, we summarize the spin- and polarization-resolved Monte-Carlo method for the Breit-Wheeler process.
\subsection*{B1: spin- and polarization-resolved pair production probability}
The pair production probability including all the polarization and spin characteristics takes the form
\begin{align}\nonumber\label{PPP2}
 dW^{P}\left(\bm{\xi},\bm{\zeta}_-,\bm{\zeta}_+\right) & =\frac{1}{2}\left(dW_{11}+dW_{22}\right)+\frac{\xi_{1}}{2}\left(dW_{11}-dW_{22}\right)\\\nonumber
 & -i\frac{\xi_{2}}{2}\left(dW_{21}-dW_{12}\right)+\frac{\xi_{3}}{2}\left(dW_{11}-dW_{22}\right)\\
 & =\frac{1}{2}\left(G_{0}+\xi_{1}G_{1}+\xi_{2}G_{2}+\xi_{3}G_{3}\right),
\end{align}
where
\begin{align*}\nonumber
G_0&  =\frac{\overline{C}_{0}}{2}d\varepsilon\Bigg\{\left\{ \int_{z_{p}}^{\infty}dx\textrm{K}_{\frac{1}{3}}\left(x\right)+\frac{\varepsilon_{+}^{2}+\varepsilon^{2}}{\varepsilon_{+}\varepsilon}\textrm{K}_{\frac{2}{3}}\left(z_{p}\right)\right\} \\\nonumber
 & +\left\{ \int_{z_{p}}^{\infty}dx\textrm{K}_{\frac{1}{3}}\left(x\right)-2\textrm{K}_{\frac{2}{3}}\left(z_{p}\right)\right\} \left(\bm{\zeta}_{-}\cdot\bm{\zeta}_{+}\right)\\\nonumber
 & +\left[\frac{\omega}{\varepsilon_{+}}\left(\bm{\zeta}_{+}\cdot\mathbf{b}\right)-\frac{\omega}{\varepsilon}\left(\bm{\zeta}_{-}\cdot\mathbf{b}\right)\right]\textrm{K}_{\frac{1}{3}}\left(z_{p}\right)\\
 & +\left\{ \frac{\varepsilon_{+}^{2}+\varepsilon^{2}}{\varepsilon\varepsilon_{+}}\int_{z_{p}}^{\infty}dx\textrm{K}_{\frac{1}{3}}\left(x\right)-\frac{\left(\varepsilon_{+}-\varepsilon\right)^{2}}{\varepsilon\varepsilon_{+}}\textrm{K}_{\frac{2}{3}}\left(z_{p}\right)\right\} \\
&\quad\left(\bm{\zeta}_{-}\cdot \mathbf{\hat{v}}\right)\left(\bm{\zeta}_{+}\cdot \mathbf{\hat{v}}\right)\Bigg\}
\end{align*}
\begin{align*}\nonumber
G_1 & =\frac{\overline{C}_{0}}{2}d\varepsilon\Bigg\{-\frac{\varepsilon_{+}^{2}-\varepsilon^{2}}{2\varepsilon_{+}\varepsilon}\textrm{K}_{\frac{2}{3}}\left(z_{p}\right)\mathbf{\hat{v}}\cdot(\bm{\zeta}_{+}\times\bm{\zeta}_{-})\\\nonumber
 & +\left[\frac{\omega}{\varepsilon}\left(\bm{\zeta}_{+}\cdot\mathbf{s}\right)-\frac{\omega}{\varepsilon_{+}}\left(\bm{\zeta}_{-}\cdot\mathbf{s}\right)\right]\textrm{K}_{\frac{1}{3}}\left(z_{p}\right)\\
 & -\frac{\omega^{2}}{2\varepsilon_{+}\varepsilon}\int_{z_{p}}^{\infty}dx\textrm{K}_{\frac{1}{3}}\left(x\right)\left\{ \left(\bm{\zeta}_{-}\cdot\mathbf{b}\right)\left(\bm{\zeta}_{+}\cdot\mathbf{s}\right)+\left(\bm{\zeta}_{-}\cdot\mathbf{s}\right)\left(\bm{\zeta}_{+}\cdot\mathbf{b}\right)\right\} \Bigg\}
\end{align*}
\begin{align*}\nonumber
G_2 & =\frac{\overline{C}_{0}}{2}d\varepsilon\Bigg\{-\frac{\omega^{2}}{2\varepsilon_{+}\varepsilon}\textrm{K}_{\frac{1}{3}}\left(z_{p}\right)\left[\mathbf{s}\cdot(\bm{\zeta}_{-}\times\bm{\zeta}_{+})\right]\\\nonumber
 & +\left(\frac{\omega}{\varepsilon_{+}}\int_{z_{p}}^{\infty}dx\textrm{K}_{\frac{1}{3}}\left(x\right)+\frac{\varepsilon_{+}^{2}-\varepsilon^{2}}{\varepsilon_{+}\varepsilon}\textrm{K}_{\frac{2}{3}}\left(z_{p}\right)\right)\left(\bm{\zeta}_{+}\cdot\mathbf{\hat{v}}\right)\\\nonumber
 & +\left(\frac{\omega}{\varepsilon}\int_{z_{p}}^{\infty}dx\textrm{K}_{\frac{1}{3}}\left(x\right)-\frac{\varepsilon_{+}^{2}-\varepsilon^{2}}{\varepsilon_{+}\varepsilon}\textrm{K}_{\frac{2}{3}}\left(z_{p}\right)\right)\left(\bm{\zeta}_{-}\cdot\mathbf{\hat{v}}\right)\\
 & -\frac{\varepsilon_{+}^{2}-\varepsilon^{2}}{2\varepsilon_{+}\varepsilon}\textrm{K}_{\frac{1}{3}}\left(z_{p}\right)\left[\left(\bm{\zeta}_{-}\cdot\mathbf{\hat{v}} \right)\left(\bm{\zeta}_{+}\cdot\mathbf{b}\right)+\left(\bm{\zeta}_{-}\cdot\mathbf{b}\right)\left(\bm{\zeta}_{+}\cdot\mathbf{\hat{v}} \right)\right]\Bigg\}.
\end{align*}
\begin{align}\nonumber \label{PRB_PHO}
G_3 & =\frac{\overline{C}_{0}}{2}d\varepsilon\Bigg\{-\textrm{K}_{\frac{2}{3}}\left(z_{p}\right)+\frac{\varepsilon_{+}^{2}+\varepsilon^{2}}{2\varepsilon_{+}\varepsilon}\textrm{K}_{\frac{2}{3}}\left(z_{p}\right)\left(\bm{\zeta}_{-}\cdot\bm{\zeta}_{+}\right)\\\nonumber
 & +\left[-\frac{\omega}{\varepsilon}\left(\bm{\zeta}_{+}\cdot\mathbf{b}\right)+\frac{\omega}{\varepsilon_{+}}\left(\bm{\zeta}_{-}\cdot\mathbf{b}\right)\right]\textrm{K}_{\frac{1}{3}}\left(z_{p}\right)\\\nonumber
 & -\frac{\left(\varepsilon_{+}-\varepsilon\right)^{2}}{2\varepsilon_{+}\varepsilon}\textrm{K}_{\frac{2}{3}}\left(z_{p}\right)\left(\bm{\zeta}_{-}\cdot\mathbf{\hat{v}}\right)\left(\bm{\zeta}_{+}\cdot\mathbf{\hat{v}}\right)\\
 & +\frac{\omega^{2}}{2\varepsilon_{+}\varepsilon}\int_{z_{p}}^{\infty}dx\textrm{K}_{\frac{1}{3}}\left(x\right)\left[\left(\bm{\zeta}_{-}\cdot\mathbf{b}\right)\left(\bm{\zeta}_{+}\cdot\mathbf{b}\right)-\left(\bm{\zeta}_{-}\cdot\mathbf{s}\right)\left(\bm{\zeta}_{+}\cdot\mathbf{s}\right)\right]\Bigg\}.
\end{align}\\
Here $\overline{C}_0=\frac{\alpha m^{2}}{\sqrt{3}\pi\omega^{2}}$, $z_p=\frac{2}{3\chi_\gamma}\frac{\omega^2}{\varepsilon_+\varepsilon_-}$ and {$\chi_\gamma=|F_{\mu\nu}k^\nu|/mF_{cr}$} controlling the magnitude of quantum effects, $\mathbf{\hat{v}}$ is the unit vector along velocity of the produced electron, $\mathbf{s}$ the unit vector along the transverse component of electron acceleration, and $\mathbf{b}=\mathbf{\hat{v}}\times \mathbf{s}$. The 3-vector $\bm{\xi}=\left(\xi_{1},\xi_{2},\xi_{3}\right)$ is the Stokes parameter of the incoming photon, $\omega$  the photon energy and $\varepsilon_{+}$ and $\varepsilon_{-}$ are the energy of the created positron and electron, respectively.\\

\noindent
\textit{Spin quantization axis for the produced electron}: After taking the sum over positron polarizations:
\begin{align*}\nonumber
 d\widetilde{W}^{p} \left(\bm{\xi},\bm{\zeta}_-\right) & =\frac{1}{2}\left(\widetilde{G}_{0}+\xi_{1}\widetilde{G}_{1}+\xi_{2}\widetilde{G}_{2}+\xi_{3}\widetilde{G}_{3}\right),
\end{align*}
\begin{align*}\nonumber
\widetilde{G}_0&  =\overline{C}_0d\varepsilon\Bigg\{ \int_{z_{p}}^{\infty}dx\textrm{K}_{\frac{1}{3}}\left(x\right)+\frac{\varepsilon_{+}^{2}+\varepsilon^{2}}{\varepsilon_{+}\varepsilon}\textrm{K}_{\frac{2}{3}}\left(z_{p}\right)
 -\frac{\omega}{\varepsilon}\left(\bm{\zeta}_{-}\cdot\mathbf{b}\right)\textrm{K}_{\frac{1}{3}}\left(z_{p}\right)\Bigg\}
\end{align*}
\begin{align*}\nonumber
\widetilde{G}_3 & =\overline{C}_0d\varepsilon\Bigg\{-\textrm{K}_{\frac{2}{3}}\left(z_{p}\right)
 +\frac{\omega}{\varepsilon_{+}}\left(\bm{\zeta}_{-}\cdot\mathbf{b}\right)\textrm{K}_{\frac{1}{3}}\left(z_{p}\right)\Bigg\}
\end{align*}
\begin{align*}\nonumber
\widetilde{G}_1 & =-\overline{C}_0d\varepsilon\frac{\omega}{\varepsilon_{+}}\left(\bm{\zeta}_{-}\cdot\mathbf{s}\right)\textrm{K}_{\frac{1}{3}}\left(z_{p}\right)
\end{align*}
\begin{align}\label{PPP3}
\widetilde{G}_2 & =\overline{C}_0d\varepsilon\Bigg\{ \left(\frac{\omega}{\varepsilon}\int_{z_{p}}^{\infty}dx\textrm{K}_{\frac{1}{3}}\left(x\right)-\frac{\varepsilon_{+}^{2}-\varepsilon^{2}}{\varepsilon_{+}\varepsilon}\textrm{K}_{\frac{2}{3}}\left(z_{p}\right)\right)\left(\bm{\zeta}_{-}\cdot\mathbf{\hat{v}} \right)\Bigg\},
\end{align}
which can be rewritten in the form
\begin{align}\nonumber
& d\widetilde{W}^{p} \left(\bm{\xi},\bm{\zeta}_-\right)=\frac{1}{2}\left( a_{-}+\bm{\zeta}_{-}\cdot\bm{b}_{-}\right)\\\nonumber
 &a_{-}=\overline{C}_{0}d\varepsilon\left[\int_{z_{p}}^{\infty}dx\textrm{K}_{\frac{1}{3}}\left(x\right)+\frac{\varepsilon_{+}^{2}+\varepsilon^{2}}{\varepsilon_{+}\varepsilon}\textrm{K}_{\frac{2}{3}}\left(z_{p}\right)-\xi_{3}\textrm{K}_{\frac{2}{3}}\left(z_{p}\right)\right]\\\nonumber
& \bm{b}_{-}=-\overline{C}_{0}d\varepsilon\left\{ \xi_{1}\frac{\omega}{\varepsilon_{+}}\mathbf{s}\textrm{K}_{\frac{1}{3}}\left(z_{p}\right)+\left(\frac{\omega}{\varepsilon}-\xi_{3}\frac{\omega}{\varepsilon_{+}}\right)\mathbf{b}\textrm{K}_{\frac{1}{3}}\left(z_{p}\right)\right.\\\nonumber
&\left.+\left[-\frac{\omega}{\varepsilon}\int_{z_{p}}^{\infty}dx\textrm{K}_{\frac{1}{3}}\left(x\right)+\frac{\varepsilon_{+}^{2}-\varepsilon^{2}}{\varepsilon_{+}\varepsilon}\textrm{K}_{\frac{2}{3}}\left(z_{p}\right)\right]\xi_{2}\mathbf{\hat{v}}\right\}.\\
\end{align}
The final polarization vector of the produced electron resulting from the scattering process itself is $\bm{\zeta}^-_f=\frac{\bm{b}_-}{a_-}$, which determines the spin quantization axis for the produced electron  $\bm{\zeta}_{f}^{-}$: $\bm{n}^-=\bm{\zeta}_f^-/|\bm{\zeta}_f^-|$.\\

\noindent
\textit{Spin quantization axis for the produced positron}: After taking the sum over electron polarizations:
\begin{align*}\nonumber
 d\overline{W}^{p} \left(\bm{\xi},\bm{\zeta}_+\right)& =\frac{1}{2}\left(\overline{G}_{0}+\xi_{1}\overline{G}_{1}+\xi_{2}\overline{G}_{2}+\xi_{3}\overline{G}_{3}\right),
\end{align*}
\begin{align*}\nonumber
\overline{G}_0&  =\overline{C}_0d\varepsilon\Bigg\{ \int_{z_{p}}^{\infty}dx\textrm{K}_{\frac{1}{3}}\left(x\right)+\frac{\varepsilon_{+}^{2}+\varepsilon^{2}}{\varepsilon_{+}\varepsilon}\textrm{K}_{\frac{2}{3}}\left(z_{p}\right)
 +\frac{\omega}{\varepsilon}_{+}\left(\bm{\zeta}_{+}\cdot\mathbf{b}\right)\textrm{K}_{\frac{1}{3}}\left(z_{p}\right)\Bigg\}
\end{align*}
\begin{align*}\nonumber
\overline{G}_3 & =\overline{C}_0d\varepsilon\Bigg\{-\textrm{K}_{\frac{2}{3}}\left(z_{p}\right)
 -\frac{\omega}{\varepsilon}\left(\bm{\zeta}_{+}\cdot\mathbf{b}\right)\textrm{K}_{\frac{1}{3}}\left(z_{p}\right)\Bigg\}
\end{align*}
\begin{align*}\nonumber
\overline{G}_1 & =\overline{C}_0d\varepsilon\frac{\omega}{\varepsilon}\left(\bm{\zeta}_{+}\cdot\mathbf{s}\right)\textrm{K}_{\frac{1}{3}}\left(z_{p}\right)
\end{align*}
\begin{align}\label{PPP3}
\overline{G}_2 & =\overline{C}_0d\varepsilon\Bigg\{ \left(\frac{\omega}{\varepsilon}_{+}\int_{z_{p}}^{\infty}dx\textrm{K}_{\frac{1}{3}}\left(x\right)+\frac{\varepsilon_{+}^{2}-\varepsilon^{2}}{\varepsilon_{+}\varepsilon}\textrm{K}_{\frac{2}{3}}\left(z_{p}\right)\right)\left(\bm{\zeta}_{+}\cdot\mathbf{\hat{v}} \right)\Bigg\},
\end{align}
which can also be written as
\begin{align}\nonumber
 &d\overline{W}^{p} \left(\bm{\xi},\bm{\zeta}_+\right)=\frac{1}{2}\left(a_{+}+\bm{\zeta}_{+}\cdot \bm{b}_{+}\right)\\\nonumber
&a_{+}=\overline{C}_{0}d\varepsilon\Bigg\{\int_{z_{p}}^{\infty}dx\textrm{K}_{\frac{1}{3}}\left(x\right)+\frac{\varepsilon_{+}^{2}+\varepsilon^{2}}{\varepsilon_{+}\varepsilon}\textrm{K}_{\frac{2}{3}}\left(z_{p}\right)-\xi_{3}\textrm{K}_{\frac{2}{3}}\left(z_{p}\right)\Bigg\}\\\nonumber
&\bm{b}_{+}=\overline{C}_{0}d\varepsilon\left\{ \xi_{1}\textrm{K}_{\frac{1}{3}}\left(z_{p}\right)\frac{\omega}{\varepsilon}\mathbf{s}+\left(\frac{\omega}{\varepsilon}_{+}-\xi_{3}\overline{C}_{0}d\varepsilon\frac{\omega}{\varepsilon}\right)\mathbf{b}\textrm{K}_{\frac{1}{3}}\left(z_{p}\right)\right.\\ \nonumber
&\left.+\xi_{2}\mathbf{\hat{v}}\left(\frac{\omega}{\varepsilon}_{+}\int_{z_{p}}^{\infty}dx\textrm{K}_{\frac{1}{3}}\left(x\right)+\frac{\varepsilon_{+}^{2}-\varepsilon^{2}}{\varepsilon_{+}\varepsilon}\textrm{K}_{\frac{2}{3}}\left(z_{p}\right)\right)\right\}.\\
\end{align}
The final polarization vector of the produced positron resulting from the scattering process itself is $\bm{\zeta}^+_f=\frac{\bm{b}_+}{a_+}$, which determines the spin quantization axis for the produced positron: $\bm{n}^+=\bm{\zeta}^+_f/|\bm{\zeta}^+_f|$.

After taking the sum over positron and electron polarizations, we get the spin unresolved pair production probability:
\begin{align}\label{WP0}
dW_{T}^{P}&\left(\bm{\xi}\right)=a_+.
\end{align}
\subsection*{ B2: Polarization-resolved no-production probability}
If a pair production event is rejected, the photon polarization  should also change due to the dependency of no-pair-production probability on photon polarization:
\begin{align}\nonumber \label{PRB_NP}
&W^{NP}\left(\boldsymbol{\xi},\boldsymbol{\xi}'\right)=\frac{1}{2}\left(c^{NP}+\boldsymbol{d}^{NP}\cdot\boldsymbol{\xi}'\right)\\\nonumber
&c^{NP}=1-\int_{0}^{\omega}a_{+}d\varepsilon\Delta t\\\nonumber
&\bm{d}^{NP}=\boldsymbol{\xi}\left(1-\int_{0}^{\omega}d\varepsilon\overline{C}_{0}\left[\int_{z_{p}}^{\infty}dx\textrm{K}_{\frac{1}{3}}\left(x\right)+\frac{\varepsilon_{+}^{2}+\varepsilon^{2}}{\varepsilon\varepsilon_{+}}\textrm{K}_{\frac{2}{3}}\left(z_{p}\right)\right]\Delta t\right)\\
&+\int_{0}^{\omega}d\varepsilon\overline{C}_{0}\hat{e}_{3}\textrm{K}_{\frac{2}{3}}\left(z_{p}\right)\Delta t.
\end{align}
where $\hat{e}_3=(0,0,1)$. The final polarization state of the photon after the no-pair-production step becomes $\bm{\xi}_f^{NP}=\bm{d}^{NP}/c^{NP}$, which defines a quantization axis for photon polarization: $\bm{n}^{NP}=\bm{\xi}_f^{NP}/|\bm{\xi}_f^{NP}|$.

The polarization induced by the no-pair production process can be estimated as
\begin{align}\nonumber\label{dXI_NP}
\Delta\bm{\xi}^{NP}&=W^{NR}\left(\bm{\xi}_{f}^{NP}-\bm{\xi}_{i}\right)\\
&=\int_{0}^{\omega}d\varepsilon\overline{C}_{0} \left(\hat{e}_{3}-\left(\boldsymbol{\xi}_{i}\cdot\hat{e}_{3}\right)\boldsymbol{\xi}_{i}\right)\textrm{K}_{\frac{2}{3}}\left(z_{p}\right)\Delta t.
\end{align}
In our scheme, the intermediate gamma photons are linearly polarized with $\xi_3\approx\pm1$, such that the polarization of the photon is unchanged during the no-pair production process, i.e. $\Delta\xi^{NP}\approx0$. Therefore, no-pair-production polarization  is trivial in this study.

\subsection*{B3: Algorithm of event generation}
1. \textit{Decide pair production event}: At each simulation step, the pair production and the electron energy are determined by the probability of Eq. (\ref{WP0}), using the common stochastic procedure.

(1) Generate two random numbers $r_1,r_2\in [0,1]$ with uniform probability.

(2) Compute the pair production probability $P(r_1)=dW_T^P(\bm{\xi}, r_1\omega)\Delta t$  for the given initial photon polarization $\bm{\xi}$,  electron energy  $\varepsilon=r_1\omega$ and positron energy $\varepsilon_+=(1-r_1)\omega$.

(3) If $r_2<P(r_1)$, an $e^+e^-$ pair is created. Otherwise, reject.\\

2. \textit{Decide the polarization of outgoing particles}: \\

\textbf{Case 1}: $P(r_1)>r_2$: pair production occurs. After each pair production, the spin of the produced electron (positron) is either parallel or antiparallel to $\bm{n}^-$ ($\bm{n}^+$)  using the stochastic procedure with another random number $r_3\in[0,1]$. With the given $\varepsilon_-$, $\varepsilon_+$ and photon polarization $\bm{\xi}$, compute the pair production probability $P_{\bm{\zeta_-}\bm{\zeta_+}}=dW^P(\bm{\xi},\bm{\zeta_-},\bm{\zeta_+})\Delta t$ with $\{\bm{\zeta_-},\bm{\zeta_+}\}\in \{\uparrow,\downarrow\}$ indicating parallel or antiparallel with respective quantization axis.

(1)  If $r_3<P_{\downarrow\downarrow}$, the electron is spin down with respect to $\bm{n}^-$ and the positron is  spin down with respect to $\bm{n}^+$, i.e. $\bm{\zeta}_-=-\bm{n}^-$, $\bm{\zeta}_+=-\bm{n}^+$

(2) If $P_{\downarrow\downarrow}<r_3<P_{\downarrow\downarrow}+P_{\downarrow\uparrow}$, $\bm{\zeta_-}=-\bm{n}^-$ and $\bm{\zeta_+}=\bm{n}^+$.

(3) If $P_{\downarrow\downarrow}+P_{\downarrow\uparrow}<r_3<P_{\downarrow\downarrow}+P_{\downarrow\uparrow}+P_{\uparrow\downarrow}$, $\bm{\zeta_-}=\bm{n}^-$ and $\bm{\zeta_+}=-\bm{n}^+$.

(4) If $P_{\downarrow\downarrow}+P_{\downarrow\uparrow}+P_{\uparrow\downarrow}<r_3<P_{\downarrow\downarrow}+P_{\downarrow\uparrow}+P_{\uparrow\downarrow}+P_{\uparrow\uparrow}$, $\bm{\zeta_-}=\bm{n}^-$ and $\bm{\zeta_+}=\bm{n}^+$.

\textbf{Case 2}: $P(r_1)<r_2$: pair production is rejected. The photon polarization state collapses into one of its basis states defined with respect to $\bm{n}^{NP}$.

(1) Generate another random number $r_4\in[0,1]$.

(2) Compute the no-pair-production probability $P_{\bm{\xi}'}=W^{NP}\left(\bm{\xi},\bm{\xi}'\right)$ for a given initial photon polarization $\bm{\xi}$. Here $\bm{\xi}'\in\{\uparrow,\downarrow\}$ indicates spin parallel or antiparallel with $\bm{n}^{NP}$.

(3) If $P_\uparrow/\left(P_\uparrow+P_\downarrow\right)>r_4$, $\bm{\xi}'=\bm{n}^{NP}$. Otherwise, $\bm{\xi}'=-\bm{n}^{NP}$.
}
\\

\bibliography{1}

\end{document}